\pdfoutput=1
\documentclass[journal]{IEEEtran}

\usepackage[utf8]{inputenc}
\usepackage[T1]{fontenc}
\usepackage{mathptmx}
\usepackage{amsmath,amssymb}
\usepackage{cite}
\usepackage{booktabs,longtable,tabularx,array}
\usepackage{graphicx}
\usepackage{xcolor}
\usepackage{enumitem}
\usepackage{float}
\usepackage{tikz}
\usetikzlibrary{arrows.meta,positioning,fit,shapes.geometric}
\usepackage{listings}
\usepackage{soul}
\usepackage{hyperref}
\usepackage[protrusion=true,expansion=false]{microtype}

\setlist{nosep,leftmargin=*}
\hypersetup{hidelinks}
\definecolor{salusblue}{HTML}{1F4E79}
\definecolor{salusgreen}{HTML}{2E7D32}
\definecolor{salusorange}{HTML}{C55A11}
\definecolor{salusgray}{HTML}{F1F4F7}
\sethlcolor{yellow!25}
\newcommand{\framework}{\textsc{SEAL-DT}}
\newcommand{\status}[1]{\textsf{#1}}
\newcommand{\NA}{\textemdash}
\newcommand{\projectrepo}{\href{https://github.com/varlamis/SALUS_mathsim}{project repository}}
\soulregister\projectrepo7
\newcommand{\projectreporef}{\projectrepo}

\newcolumntype{Y}{>{\raggedright\arraybackslash}X}
\newcolumntype{L}[1]{>{\raggedright\arraybackslash}p{#1}}

\title{\framework{}: An End-to-End Security Engineering and Assurance Framework for Cross-Domain Digital Twins}

\author{Iraklis~Varlamis, Stelios~Zindros, Christos~Sardianos, Thomas~Lagkas,
Panagiotis~Sarigiannidis, and Georgios~Th.~Papadopoulos%
\thanks{I. Varlamis, S. Zindros, C. Sardianos and G. Th. Papadopoulos are with
the Department of Informatics and Telematics, Harokopio University of Athens,
Athens, Greece (e-mail: varlamis@hua.gr; itp23106@hua.gr; sardianos@hua.gr;
g.th.papadopoulos@hua.gr).}%
\thanks{T. Lagkas is with the Department of Informatics, Democritus University
of Thrace, Kavala, Greece (e-mail: tlagkas@cs.duth.gr).}%
\thanks{P. Sarigiannidis is with the Department of Electrical and Computer
Engineering, University of Western Macedonia, Kozani, Greece (e-mail:
psarigiannidis@uowm.gr).}}

\begin{document}
\maketitle

\begin{abstract}
Cyber-physical systems connect physical processes, software, networks and
organizations. Securing these systems requires a clear account of what happened,
what was observed, which actions were authorized and what evidence supports each
decision. Digital Twin (DT) descriptions often do not explain how these elements
remain consistent. They may also blur the distinction between physical events
and uncertain observations, or leave simulation results disconnected from their
requirements and assumptions.
This paper presents \framework{} (Security Engineering and Assurance Lifecycle
for Digital Twins), a structured approach to building and
evaluating secure DTs. It brings entity definitions, process models,
deterministic simulations, data contracts and evidence records, 
automated tests and reproducible results into one connected model. 
The main contribution is an end-to-end security framework for
DTs, implemented from scenario definition and entity/process modelling through
simulation, security decision logic, evidence handling, validation and
reproducible reporting.
This makes it possible to trace security decisions and simulation results
back to their data, assumptions and requirements. In this approach, a DT is more
than a copy of system state: it is a representation that clearly states
what it can observe, how accurate and current its information is, where that
information came from and which decisions it is allowed to support.
We demonstrate the approach in five scenarios: Car2X protection, passive IoT-
device detection, an industrial control surrogate, lawful acquisition in a
smart home, and medical-IoT monitoring that preserves continuity of care. Each
implementation includes formal models, deterministic and negative experiments,
uncertainty analyses, schema-valid fixtures, technical reports and
verified manifests. Across the scenario set, multiple configurations are
tested under both normal and adverse conditions. The results show that the
framework supports consistent model construction, traceable execution and
comparable internal validation. They also demonstrate that security findings
can be linked to observable evidence. 
\end{abstract}


\section{Introduction}
\label{sec:introduction}

\IEEEPARstart{D}{igital} Twins are increasingly used to reason about systems whose physical,
computational and human parts evolve together. They combine models with
operational data to represent a system, track its changing state and support
decisions throughout its life cycle \cite{glaessgen2012}. Current standards and
industrial frameworks emphasize observable entities, synchronization,
information exchange and interoperability \cite{iso23247,dtcinterop,aas}. 
However, security-focused deployments must keep five things separate: what
actually happened, what sensors recorded, what the digital twin accepted as
valid, what an analyst is authorized to see, and what evidence supports a
conclusion. For example, a sensor may detect a device, but the twin may reject
the report and privacy rules may prevent an analyst from seeing the device's
identity; treating the detection as confirmed and visible fact would make a
simulation appear certain without enough evidence.

The contribution of \framework{} is a Security Engineering and
Assurance Lifecycle for Digital Twins: a complete framework for designing,
implementing and validating security-oriented DT scenarios across heterogeneous
cyber-physical domains. The framework represents these layers separately while
linking every system transition, security decision and simulation result to its
requirements, assumptions, data and supporting evidence. This structure makes
security conclusions auditable and prevents model predictions from being
mistaken for observations, authorized actions or validated claims.
Its implementation covers the full lifecycle
from source-governed scenario scoping to executable simulations, negative
security experiments, evidence/custody records, automated tests and
reproducible manifests. To demonstrate that the approach applies beyond a
single domain, we evaluate it through a heterogeneous set of security-DT
scenarios.
Despite their different settings and dynamics, these scenarios share the same
assurance questions:

\begin{itemize}
  \item Which physical, digital, logical, human, organizational and data
  entities exist, and which source justifies each one?
  \item Which transitions are possible, what is observable, and how do delay,
  loss, ambiguity and stale state affect the digital representation?
  \item Which detector or policy requested an action, which capability actually
  applied it, and which safety or authorization constraint overruled it?
  \item Which visible observations and evidence records support a reconstructed
  claim, and can the result be reproduced from declared inputs?
\end{itemize}

To address these questions systematically, \framework{} contributes the following elements:

\begin{enumerate}
  \item a DT definition that binds identity, referent, state, observations,
  update rules, fidelity, freshness, provenance, policy and decision authority;
  \item a source-governed case method that records scope, source precedence,
  assumptions, prohibited capabilities and integration boundaries;
  \item a canonical entity and graph model linked to process and behavioural
  views over events, queues, observation loss and twin updates;
  \item a security-governance layer for detection, authorization, safety guards,
  minimization, evidence, custody and reconstruction;
  \item an assurance package method covering experiments, uncertainty analysis,
  competency queries, tests, reports, manifests and bounded claim discipline; and
  \item a comparative application to heterogeneous scenarios, with measured synthetic
  results reported only where generated artifacts exist.
\end{enumerate}

Taken together, these contributions position the work as an
implemented security framework for DT-based cyber-physical threat scenarios,
rather than only a conceptual modelling proposal. The framework keeps physical
events, observations, accepted twin state, authorized decisions, evidence and
claims distinct, while preserving the links needed to audit and reproduce a
result. It combines this separation with source-governed assumptions, canonical
entities, executable behaviour and assurance artifacts. The evaluation also
shows how the framework can be reused without removing domain-specific dynamics
or constraints.

The remainder of the article is organized as follows. Section~\ref{sec:related-work}
reviews relevant DT, security and provenance foundations.
Section~\ref{sec:framework} defines the proposed framework as a chain from
source-governed case to assurance package. Section~\ref{sec:applications}
applies the framework to the use cases, while
Section~\ref{sec:results} presents the simulation design and comparative results.
Section~\ref{sec:limitations} discusses current limitations and future work, and
Section~\ref{sec:conclusions} summarizes the main conclusions. Finally, the
Appendix provides the detailed UC1 implementation view.

\section{Related Work and Related Frameworks}
\label{sec:related-work}

To position the proposed framework, this section reviews three complementary
strands of prior work. First, Digital Twin definitions and reference
architectures establish how physical or logical referents are represented and
kept aligned with their digital counterparts. Second, cyber-physical security
and model-assurance research provides the concepts needed to reason about
trustworthiness, uncertainty and operational risk. Third, process, schema and
provenance standards support interoperable execution and auditable evidence.
Taken together, these three strands identify the architectural, assurance and
provenance foundations adopted by \framework{}, while also exposing the gap that
motivates the framework chain introduced in Section~\ref{sec:framework}: moving
from source-governed case definition to a reviewable assurance package.

\subsection{Digital Twin foundations and reference architectures}

Digital twin research and standardization have converged over the years around a set of
foundational definitions, reference architectures and interoperability
principles.
The original aerospace formulation by Glaessgen and Stargel, developed in the
NASA and U.S. Air Force context, established the central idea of a digital
representation informed by onboard and historical data and used to reason about
the lifecycle of a corresponding physical system \cite{glaessgen2012}.
Respectively, ISO 23247 provided terms, principles and a reference
architecture for manufacturing DTs, including observable manufacturing
elements and information exchange \cite{iso23247}. In a similar spirit,
the Asset Administration
Shell (AAS) specifies a metamodel, submodels, identifiers and exchange mappings
for interoperable representations of industrial assets \cite{aas}, and the Digital
Twin Consortium's interoperability framework emphasizes implementation-agnostic,
secure and scalable interoperation among independently operable systems
\cite{dtcinterop}.

Together, these frameworks establish the architectural and semantic baseline for
DT work: how represented entities are identified and structured, how observations
and asset information are exchanged, and how interoperability, lifecycle scope
and fidelity are framed. \framework{} builds on this baseline, but targets a
security-specific problem: a security claim is useful only when its origin, input
data, reasoning steps, supporting evidence and uncertainty can be checked. It
therefore links each protected entity to the security observations that describe
its state, the assessment or control decision derived from them, the evidence
supporting that decision and the confidence attached to the result. This makes a
DT usable for security work, where analysts must be able to inspect why an alert,
risk score or control recommendation was produced and what evidence bounds the
resulting claim. The same structure can be used for physical assets, logical
services and security hypotheses, as long as each twin clearly states what it
refers to, what it may observe and what level of fidelity it claims.

\subsection{Cyber-physical systems, security and model assurance}

The NIST CPS Framework treats cyber-physical systems as interacting physical,
digital and human components and organizes analysis through conceptualization,
realization and assurance facets, alongside concerns such as timing, data,
trustworthiness, boundaries and lifecycle \cite{nistcps}. This broad systems
view motivates the cross-domain scope of \framework{}. Recent CPS security
surveys further show that attacks and defenses must be analyzed across the
physical, cyber and cyber-physical domains, since a software or network
compromise can become a physical safety or process-integrity problem
\cite{yu2023cpssecurity}. Recent DT surveys make the same point from the twin
side: DT security depends on the architectural layer being protected, the
application domain, the modeling technique and the operating mode
\cite{jeremiah2024dtsurvey,qureshi2025securitydt}.

Work on DT-based cyber-attack detection in manufacturing demonstrates how
physics-based, data-driven and expert models can be combined to distinguish
expected anomalies from attacks \cite{balta2023}. CRYSTAL brings this idea
closer to assurance by combining attack models, formal verification, runtime
monitoring and a compact ``Tiny Digital Twin'' reference model for detecting CPS
attacks at runtime \cite{moradi2024crystal}. \framework{} generalizes
these patterns to detection, policy, safety, evidence and reconstruction, while
requiring each operational inference to remain within its observation boundary,
declared authority and stated uncertainty.

The closest prior work is the security-enhancing DT literature, which treats a
DT not only as an operational mirror but also as a security instrument.
Prior work shows DTs supporting CPS security design, testing, intrusion
detection and SOC/SIEM analysis for operational technology
\cite{eckhart2019cpssecurity,dietz2020soc}. Security-enhancing DT surveys
further define the characteristics of twins that actively improve security and
map the area across models, tools, domains and operating modes
\cite{eckhart2023sedt,qureshi2025securitydt}. Recent industrial DT-security
work is especially relevant because it treats twins as infrastructure for
incident detection and response in IIoT, manufacturing and energy settings
\cite{katsikas2026industrialdt}. \framework{} follows this positive direction,
but makes the security claim more explicit by binding observations, decisions,
policy actions and reconstruction claims to typed evidence and provenance.
Industrial DT forensics is another close line of work. Replication-based ICS
twins emphasize state replication, replay and fidelity for post-incident
analysis, where the physical plant cannot always be imaged or disturbed
\cite{dietz2021icsforensics}. ATT\&CK-oriented DT work also shows how
synchronized ICS state changes can become evidence-preserving descriptions and
typed dependency graphs \cite{katsikas2026attckdt}. \framework{} shares this
forensic-readiness motivation, but extends it into an end-to-end evidence
contract: the twin records what was observable, how observations were derived,
which claims they support and which uncertainty or custody limits prevent
stronger conclusions.
Model credibility and verification/validation work also shapes the assurance
discipline. NIST, ASME V\&V 40 and NASA-STD-7009 all treat credibility as tied
to intended use, uncertainty, input pedigree, verification, validation and
reporting \cite{nistdtcredibility,asmevv40,nasastd7009}. \framework{} imports
this stance into security DTs by separating executable behaviour from claim
qualification: a run can complete deterministically while its assurance package
still states limited fidelity, synthetic data scope, missing calibration or
insufficient authority for an operational claim.
Runtime assurance provides the complementary control perspective. The Simplex
architecture separates an advanced controller from a trusted safety controller
and switches authority when monitored conditions leave the certified envelope
\cite{sha1996simplex}. Shielding and runtime-enforcement work similarly places
a monitor around a reactive component so that critical properties are enforced
online even when the main controller is too complex to fully verify
\cite{bloem2015shield}. \framework{} uses the same separation principle:
DT-derived detections, policies and mitigations must state their authority
boundary, and safety- or security-relevant actions are represented as
controlled transitions with evidence and failure modes.
\subsection{Process, schema and provenance standards}

Security assurance requires processes, schemas and provenance records that make
security artifacts both interpretable by analysts and checkable by tools.
Several standards illustrate this need. For example,
NIST SP 800-61 frames incident response as part of cybersecurity risk
management, linking preparation, detection, response and recovery to the NIST
Cybersecurity Framework 2.0 \cite{nistir2025}. NIST OSCAL represents control
catalogs, baselines, system security plans, assessments and assessment results
in machine-readable formats \cite{oscal}. For operational data, the Open
Cybersecurity Schema Framework (OCSF) defines a common schema for security
events, findings and observables \cite{ocsf}. STIX 2.1 and TAXII 2.1 support
structured cyber threat intelligence and its exchange \cite{stix21,taxii21},
while CSAF 2.0 supports machine-readable security advisories about products,
vulnerabilities, impact and remediation \cite{csaf20}.

\framework{} follows the same direction by treating security evidence as
explicit typed artifacts rather than informal notes. JSON Schema Draft 2020-12
supplies machine-validatable structure and constraints for JSON artifacts
\cite{jsonschema}. W3C PROV models entities, activities, agents and derivations
\cite{prov}. NIST SP 800-86 separates collection, examination, analysis and
reporting in digital forensics and emphasizes preservation and documented
handling \cite{nistforensics}. BPMN 2.0.2 defines a notation, metamodel and
interchange format for business processes and collaborations \cite{bpmn};
\framework{} uses it only as a responsibility and message-flow view, while
schemas, configuration and code retain the executable behavioural semantics. In
this way, the framework combines process structure, schema validation and
provenance through source records, evidence objects, derivation links and
custody chains; it does not claim that a local hash chain alone makes evidence
legally admissible.

Assurance-case standards clarify how evidence should be turned into an
argument. Goal Structuring Notation (GSN) provides a graphical structure for
claims, strategies, context, assumptions, justifications and evidence
\cite{gsnstandard}. The OMG Structured Assurance Case Metamodel (SACM) provides
a metamodel for machine-processable assurance cases and related artifact
references \cite{sacm20}. \framework{} does not require authors to use a
specific assurance-case editor, but it follows the same claim--argument--
evidence logic: every reported result is treated as a qualified claim whose
supporting observations, tests, assumptions and residual uncertainty must be
available for review.
Forensic-readiness standards add the evidence-handling boundary. ISO/IEC 27037
focuses on identification, collection, acquisition and preservation of
potential digital evidence, while ISO/IEC 27043 provides incident-investigation
principles and process models for scenarios involving digital evidence
\cite{iso27037,iso27043}. These standards reinforce a key distinction in
\framework{}: the DT can prepare, preserve and explain evidence-bearing
artifacts, but admissibility and investigative sufficiency depend on broader
organizational, legal and procedural controls. The framework therefore records
custody, minimization, hashes, source precedence and derivation, while avoiding
the stronger claim that technical traceability alone establishes forensic
validity.
\begin{table*}[t]
\centering
\scriptsize
\setlength{\tabcolsep}{1pt}
\caption{Comparison of related security-DT frameworks and the proposed framework.}
\label{tab:related-comparison}
\begin{tabularx}{\textwidth}{@{}L{27mm}>{\centering\arraybackslash}p{12mm}>{\centering\arraybackslash}p{18mm}>{\centering\arraybackslash}p{17mm}>{\centering\arraybackslash}p{15mm}>{\centering\arraybackslash}p{19mm}>{\centering\arraybackslash}p{18mm}L{32mm}@{}}
\toprule
Framework & Twin model & Monitoring / detection & Trust / provenance &
Incident response & Formal / runtime assurance & End-to-end assurance chain &
Main limitation relative to \framework{} \\
\midrule
CPS cybersecurity DT \cite{zhao2022dtcyber}
& $\checkmark$ & $\checkmark$ & $\sim$ & $\sim$ & \NA{} & \NA{}
& Focuses on CPS cybersecurity analysis, but does not define a full typed
evidence chain from source declaration to qualified result. \\
Trusted Twins \cite{suhail2022trustedtwins}
& $\checkmark$ & $\sim$ & $\checkmark$ & \NA{} & \NA{} & \NA{}
& Strengthens trust and integrity through blockchain, but does not cover
process-level assurance, decision authority and reconstruction together. \\
SEDT incident-response framework \cite{suhail2025sedtir}
& $\checkmark$ & $\sim$ & $\sim$ & $\checkmark$ & \NA{} & $\sim$
& Covers the SEDT lifecycle for incident response, but does not formalize a
cross-use-case executable evidence and uncertainty contract. \\
Cybersecurity DT blueprint \cite{repetto2026cdt}
& $\checkmark$ & $\checkmark$ & $\sim$ & $\checkmark$ & $\sim$ & $\sim$
& Addresses federated security operations across service chains, but leaves the
artifact-level assurance chain and validation contract less explicit. \\
\framework{}
& $\checkmark$ & $\checkmark$ & $\checkmark$ & $\checkmark$ & $\checkmark$
& $\checkmark$
& Integrates entity modeling, observations, policy, evidence, provenance,
uncertainty and reconstruction in one executable assurance discipline. \\
\bottomrule
\end{tabularx}
\end{table*}

\subsection{Gap addressed}

The related work shows several mature but partial answers to the security DT
problem. ISO 23247, AAS and the Digital Twin Consortium define how twins,
industrial assets and interoperable DT systems can be structured
\cite{iso23247,aas,dtcinterop}. DT-based detection frameworks, including
manufacturing attack detection and CRYSTAL, show how model-based, data-driven
and formal techniques can support CPS attack detection and runtime assurance
\cite{balta2023,moradi2024crystal}.
Security-enhancing DT work shows how twins can support security
analysis, SOC simulation, intrusion detection, incident response and ICS
forensic replay
\cite{eckhart2019cpssecurity,dietz2020soc,eckhart2023sedt,dietz2021icsforensics,suhail2025sedtir,qureshi2025securitydt,katsikas2026industrialdt,katsikas2026attckdt}.
Process and evidence standards
such as OSCAL, OCSF, STIX/TAXII, CSAF, JSON Schema and PROV define useful
building blocks for controls, events, advisories, validation and provenance
\cite{oscal,ocsf,stix21,taxii21,csaf20,jsonschema,prov}.
Credibility, runtime-assurance, assurance-case and
forensic-readiness standards add discipline for deciding when a DT result can
support a security or safety-relevant claim
\cite{nistdtcredibility,asmevv40,nasastd7009,sha1996simplex,bloem2015shield,gsnstandard,sacm20,iso27037,iso27043}.

Several works move toward such composition. CPS cybersecurity DT frameworks link
the physical system, cyber representation and security analysis tasks, with
emphasis on monitoring and attack analysis \cite{zhao2022dtcyber}. Trusted
Twins combine blockchain-backed trust mechanisms with DT-based CPS situational
awareness, strengthening integrity and trust in the twinning process
\cite{suhail2022trustedtwins}. Security-enhancing DT lifecycle work for
incident response covers design, deployment, operation and end-of-life
considerations \cite{suhail2025sedtir}. Cybersecurity DT blueprints for
federated, multi-ownership digital service chains emphasize cross-domain
cooperation, threat hunting and adaptive response \cite{repetto2026cdt}.

Table~\ref{tab:related-comparison} summarizes how these works relate to the
building blocks used by \framework{}. The comparison is qualitative:
$\checkmark$ denotes explicit support, $\sim$ denotes partial or domain-specific
support, and \NA{} denotes that the feature is not a main concern of the cited
work.

What remains less well supported in the related work is the composition of these
concerns into one testable assurance chain. A security DT does not only need an
asset model, an event schema, a detection model or a provenance record in
isolation. It needs a way to keep source precedence, assumptions, entity
identifiers, observation boundaries, process views, policy limits, evidence
objects and uncertainty statements consistent from case definition to executable
result and claim qualification. Without that chain, a twin may produce alerts or
recommendations that are difficult to audit, compare across use cases, reproduce
or challenge.

This gap is especially visible when the closest prior work is considered
together. Security-enhancing DTs, model credibility, runtime assurance,
assurance cases and forensic readiness each address part of the problem, but
they are usually applied separately. \framework{} contributes the connective
tissue: source-governed scenario definition, typed entities, executable
simulations, schema-validated evidence, provenance, uncertainty statements,
policy authority and reconstruction claims are kept in one reviewable package.
\framework{} addresses this gap as a composition and assurance discipline, not
as a replacement for the standards above. Its benefit is to make security DT
claims explicit, executable and inspectable: inputs are typed, transitions are
controlled, decisions remain within declared authority, evidence is linked to
its sources and results carry provenance and uncertainty. This allows the same
framework to support detection, policy assessment, safety reasoning and
forensic reconstruction while preserving a clear boundary between what the twin
observes, what it infers, what it is allowed to decide and what the assurance
package is strong enough to claim.

Compared with the surveyed literature, four elements appear most distinctive.
\framework{} keeps five layers separate--physical, observed, accepted-twin,
decision/action and evidence/claim state--and applies a no-leakage rule across
them. It also uses a richer DT definition than a synchronized replica, requiring
the twin to declare its referent, observation boundary, fidelity, provenance,
uncertainty and decision authority. Requested and applied security actions are
separated by one guard pattern reused across the five domains. Finally, claim
discipline and evidence-maturity labels state whether a result is supported by
executable evidence, synthetic validation, partial calibration, expert
assumption or an open obligation. Together these elements turn DT security
functionality into a reviewable assurance chain.
\section{The \framework{} Framework for Digital-Twin Security Threat Scenarios}
\label{sec:framework}

\framework{} is an engineering framework for turning a broad cyber-physical
security scenario into an executable, traceable and reviewable Digital Twin
package. It covers different application scenarios without forcing them into
the same domain model or threat pattern. In each case the starting point is a
narrative problem statement. The end point is a reviewable package containing
source decisions, assumptions, canonical entities, process models, schemas,
executable runs, evidence objects, graph queries, tests, results and a hash
manifest.

The framework therefore treats a DT as more than a synchronized model of an
asset. A secure and verifiable DT in \framework{} is a declared representation
with an identifier $id_i$, a referent $r_i$, an accepted digital state
$\hat x_i$, admissible observations $\mathcal O_i$, an update rule $U_i$,
declared fidelity $\Phi_i$, freshness $F_i$, provenance and uncertainty
$\Delta_i$, policy constraints $P_i$ and decision authority $A_i$:
\begin{equation}
\mathcal T_i=(id_i,r_i,\hat x_i,\mathcal O_i,U_i,\Phi_i,F_i,\Delta_i,P_i,A_i).
\label{eq:dt}
\end{equation}
This compact tuple formally defines
the questions that had to be answered for each implementation: what is being
twinned, which observations may update it, how stale or uncertain the state may
be, which decisions it may influence, and which evidence supports each claim.

\subsection{Source-Governed Case to Assurance Package}

\begin{figure}[ht]
\centering
\begin{tikzpicture}[
  scale=.94,
  transform shape,
  node distance=5mm,
  box/.style={draw=salusblue,rounded corners,fill=salusgray,text width=.88\linewidth,
  minimum height=10mm,align=center,font=\small},
  arr/.style={-{Latex[length=2mm]},thick,salusblue}]
\node[box] (s) {\textbf{1. Source-governed case}: scope, source precedence,
assumptions, prohibited capabilities, integration boundary};
\node[box,below=of s] (c) {\textbf{2. Canonical model}: reusable entities,
typed relationships, zones, roles and schema references};
\node[box,below=of c] (p) {\textbf{3. Process and behaviour}: BPMN/lifecycle
views, state variables, events, queues, observation loss and twin updates};
\node[box,below=of p] (g) {\textbf{4. Security governance}: detection,
authorization, safety guards, minimization, evidence, custody and reconstruction};
\node[box,below=of g] (a) {\textbf{5. Assurance package}: experiment matrix,
Monte Carlo design, runtime graph, competency queries, tests, reports and manifest};
\draw[arr] (s)--(c); \draw[arr] (c)--(p); \draw[arr] (p)--(g); \draw[arr] (g)--(a);
\draw[arr,dashed] ([xshift=-3mm]a.west) -- ++(-8mm,0) |- ([xshift=-3mm]s.west);
\end{tikzpicture}
\caption{The \framework{} chain from source intent to validated result. The
dashed arrow shows how failed checks lead to revisions.}
\label{fig:framework}
\end{figure}
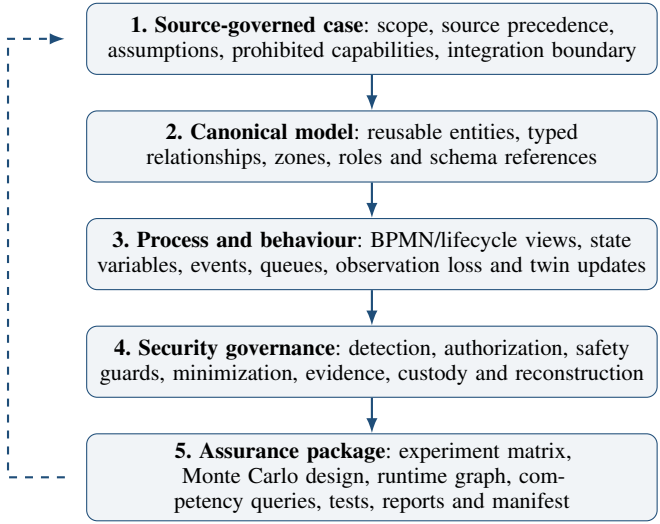

Figure~\ref{fig:framework} presents the framework as a layered assurance chain,
linking identifiers, assumptions and source references at the top to executable
results and the final report at the bottom, while keeping each intermediate
component traceable and visible.

The same chain can be instantiated for different security-threat scenarios. In a
Car2X attack setting, it can separate malicious or untrusted message validation
from the authority to influence traffic-control decisions. In smart-home
forensics, it can keep incident truth, observations, lawful collection, custody
and claims distinct.

\subsection{Source-Governed Case}
\label{sec:source-governed-case}

The source-governed case layer reduces the broad problem to one bounded
scenario with explicit scope, source precedence, assumptions, prohibited
capabilities and integration boundary. This is where \framework{} prevents two
common failures: modelling everything too vaguely, or implementing a narrow
simulator whose assumptions are invisible. For each scenario, the boundary
records:
\begin{itemize}
  \item the scope of the protected or investigated referents, such as vehicle
  messages and roadside-unit requests;
  \item the source precedence among declared scenario requirements, roadmap
  obligations, user decisions, engineering assumptions and TBD values;
  \item the assumptions, affected artifacts, default values, owners,
  replacement paths and impacts;
  \item the actions explicitly prohibited in the synthetic model, such as live
  RF transmission or real traffic-control actuation; and
  \item the integration boundary for the services allowed to observe, decide,
  collect or share Car2X evidence.
\end{itemize}

The output is a scenario contract rather than an informal description. The
contract does not eliminate uncertainty. It makes uncertainty operational by registering
assumptions with affected artifacts, default values, owner, replacement path and
impact. This is what allows a scenario package to state both what the synthetic
model showed and what it did not show.

\begin{quote}\itshape
Example scenario contract: the Car2X simulation models a synthetic test vehicle
that sends a roadside-unit request to influence a traffic-control decision during
a bounded run of $T_{\max}$ seconds. The security scenario is a forged, replayed or
stale request that attempts to pass trust validation and affect the traffic
controller. The in-scope referents are the vehicle message, the roadside-unit
request, the gateway decision and the traffic-controller command interface. The
simulation records any missing Car2X profile, timing parameter or trust threshold
as assumption $Y$ with an owner, replacement source and affected artifacts. The
run must not transmit live RF, connect to a real roadside unit or actuate a real
traffic controller. The integration boundary permits only synthetic observations,
policy decisions, evidence records and reports. A successful run must show which
request was accepted or rejected, which authority allowed or blocked the
decision, which evidence supports the result and which assumptions still limit
the claim.
\end{quote}

\subsection{Canonical Model}

The canonical model layer defines reusable entities and their typed
relationships. Zones, roles, data objects and schema references are recorded as
entity attributes where needed. Every relevant object is named as an entity and
receives a stable identifier before it appears in a process, equation, event
stream, evidence package or report. The inventory of reusable entities is:
\begin{equation}
\mathcal I=\{e_i\mid e_i=(id_i,\tau_i,name_i,src_i,z_i,\rho_i,\sigma_i,\ell_i,\theta_i)\}.
\label{eq:inventory}
\end{equation}
Here $id_i$ is the stable identifier, $\tau_i$ is the entity type, $name_i$ is
the human-readable name, $src_i$ is the source or assumption status, $z_i$ is
the zone or boundary, $\rho_i$ is the role, $\sigma_i$ is the sensitivity label,
$\ell_i$ is the lifecycle state and $\theta_i$ is an optional twin reference.
The same entity may appear in several views, but it must keep the same
identifier.

This rule is what allows \framework{} to cover different domains without
erasing their differences. A roadside unit and a traffic controller, for
example, are domain-specific objects. In \framework{}, each domain-specific
object is still represented as an entity with source status, sensitivity, zone,
lifecycle state, relationships and optional twin reference. Relationships are typed because
``observes'', ``controls'', ``mirrors'', ``authorizes'', ``supports'' and
``shares'' have different security meanings.

The canonical graph has two parts. The design graph records the declared
entities and allowed relationships:
\begin{equation}
G_0=(V_0,E_0), \qquad V_0=\mathcal I,\qquad
E_0\subseteq V_0\times R\times V_0 .
\label{eq:design-graph}
\end{equation}
In this notation, $V_0$ is the set of reusable design-time entities, $E_0$ is
the set of typed design-time relationships and $R$ is the allowed relationship
vocabulary, such as \textsf{sends}, \textsf{observes}, \textsf{validates},
\textsf{forwards}, \textsf{authorizes}, \textsf{controls} and \textsf{supports}.
Runtime execution
adds observations, alerts, decisions, evidence, custody events and claims:
\begin{equation}
G(t)=(V_0\cup V_r(t),E_0\cup E_r(t)).
\label{eq:graph}
\end{equation}
Here $V_0$ is the stable entity inventory, $V_r(t)$ is the set of runtime
vertices accumulated up to time $t$, and $E_r(t)$ is the set of runtime edges
linking those vertices to design-time entities and to each other. This graph can
support competency queries such as: which evidence supports
this claim, which authorization allowed this collection, which twin state was
stale, which source assumption affected the metric, and which process step
generated the decision.

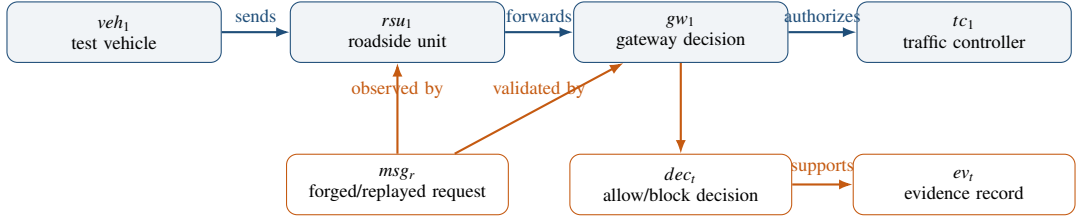
\begin{figure*}[t]
\centering
\begin{tikzpicture}[
  node distance=12mm and 9mm,
  ent/.style={draw=salusblue,rounded corners,fill=salusgray,text width=26mm,
  minimum height=8mm,align=center,font=\scriptsize},
  run/.style={draw=salusorange,rounded corners,fill=white,text width=27mm,
  minimum height=8mm,align=center,font=\scriptsize},
  edge/.style={-{Latex[length=1.8mm]},thick,salusblue},
  redge/.style={-{Latex[length=1.8mm]},thick,salusorange}]
\node[ent] (veh) {$veh_1$\\test vehicle};
\node[ent,right=of veh] (rsu) {$rsu_1$\\roadside unit};
\node[ent,right=of rsu] (gw) {$gw_1$\\gateway decision};
\node[ent,right=of gw] (tc) {$tc_1$\\traffic controller};
\node[run,below=of rsu] (msg) {$msg_r$\\forged/replayed request};
\node[run,below=of gw] (dec) {$dec_t$\\allow/block decision};
\node[run,below=of tc] (ev) {$ev_t$\\evidence record};
\draw[edge] (veh)--node[above,font=\scriptsize]{sends}(rsu);
\draw[edge] (rsu)--node[above,font=\scriptsize]{forwards}(gw);
\draw[edge] (gw)--node[above,font=\scriptsize]{authorizes}(tc);
\draw[redge] (msg)--node[above,font=\scriptsize]{observed by}(rsu);
\draw[redge] (msg)--node[above,font=\scriptsize]{validated by}(gw);
\draw[redge] (gw)--(dec);
\draw[redge] (dec)--node[above,font=\scriptsize]{supports}(ev);
\end{tikzpicture}
\caption{Example canonical Car2X graph for the Section~\ref{sec:source-governed-case}
simulation contract. Blue edges are reusable design relationships; orange edges
are runtime relationships created by the security test.}
\label{fig:car2x-canonical-graph}
\end{figure*}

For example, in the Car2X scenario, a
minimal canonical inventory may define reusable entities such as:
\begin{equation}
\begin{aligned}
e_1&=(veh_1,\textsf{physical},\text{test vehicle},Y,\\
&\quad\textsf{test-zone},\textsf{sender},\textsf{restricted},
\textsf{active},dt_{veh}),\\
e_2&=(rsu_1,\textsf{cyber},\text{roadside unit},Y,\\
&\quad\textsf{edge-zone},\textsf{receiver},\textsf{restricted},
\textsf{simulated},dt_{rsu}),\\
e_3&=(gw_1,\textsf{logical},\text{gateway decision},Y,\\
&\quad\textsf{control-zone},\textsf{validator},\textsf{internal},
\textsf{simulated},dt_{gw}),\\
e_4&=(tc_1,\textsf{cyberphysical},\text{traffic controller},Y,\\
&\quad\textsf{control-zone},\textsf{actuator},\textsf{restricted},
\textsf{simulated},dt_{tc}).
\end{aligned}
\label{eq:car2x-entities}
\end{equation}
In these tuples, $veh_1$, $rsu_1$, $gw_1$ and $tc_1$ are stable entity
identifiers; $Y$ denotes the assumption record introduced by the scenario
contract; and $dt_{veh}$, $dt_{rsu}$, $dt_{gw}$ and $dt_{tc}$ are optional twin
references for the corresponding entities. The roadside unit is typed as
\textsf{cyber} because it handles the digital request path, while the traffic
controller is typed as \textsf{cyberphysical} because its command interface is
coupled to the traffic-control process, even though real actuation is excluded
from the simulation.
The corresponding design graph can be written as:
\begin{equation}
\begin{aligned}
E_0=\{&(veh_1,\textsf{sends},rsu_1),\\
&(rsu_1,\textsf{forwards},gw_1),\\
&(gw_1,\textsf{authorizes},tc_1)\}.
\end{aligned}
\label{eq:car2x-graph}
\end{equation}
The security test then adds runtime vertices for a forged, replayed or stale
request, its validation result, the allow/block decision and the evidence record
supporting the final claim. In Figure~\ref{fig:car2x-canonical-graph},
$msg_r\in V_r(t)$ denotes the runtime request record, where $r$ indexes the
request instance; $dec_t\in V_r(t)$ denotes the decision created at time $t$;
and $ev_t\in V_r(t)$ denotes the evidence record created at time $t$. The orange
runtime relationships \textsf{observed by}, \textsf{validated by} and
\textsf{supports} are elements of $E_r(t)$. Figure~\ref{fig:car2x-canonical-graph}
illustrates the same example as a compact graph.

\subsection{Process and Behaviour Views}

The process and behaviour layer contains BPMN/lifecycle views, state variables,
events, queues, observation loss and twin updates. It defines process
responsibilities and behavioural views over the same reusable entities used by
the canonical model. Scenario packages may contain BPMN and lifecycle
artifacts, but \framework{} does not treat BPMN as the source of executable
truth. Their role is to show
responsibility, message flow, denial paths and handoff points. The executable
semantics remain in schemas, configuration and code.

The reusable process-family vocabulary is:
\begin{equation}
\begin{aligned}
\mathcal F=\{&\textsf{nominal},\textsf{attack},\textsf{detection},\\
&\textsf{policy},\textsf{evidence},\textsf{lifecycle}\}.
\end{aligned}
\label{eq:process-families}
\end{equation}
These families separate normal operation, adversarial or abnormal behaviour,
security assessment, governed response, evidence handling and recovery or
state-change management. This vocabulary covers a wide range of simulation
scenarios because every process step must either establish the baseline,
introduce or observe a threat, assess it, decide a governed response, preserve
supporting evidence or move the scenario into a later lifecycle state. It is not
intended as a universal process taxonomy; domain-specific substeps can be added
under one of these families without changing the framework structure.
For a scenario, the process view is an inventory:
\begin{equation}
\mathcal P=\{p_j\mid p_j=(pid_j,f_j,W_j,pre_j,post_j,in_j,out_j,deny_j)\}.
\label{eq:process-inventory}
\end{equation}
Here $pid_j$ is the process identifier, $f_j\in\mathcal F$ is the process
family, $W_j\subseteq V_0$ is the set of responsible entities or roles,
$pre_j$ and $post_j$ are the entry and exit conditions, $in_j$ and $out_j$ are
the consumed and produced messages or artifacts, and $deny_j$ records the
explicit denial or failure branch. This notation keeps the process view focused
on who observes, reviews, mitigates, collects or shares, while the executable
model decides what actually happens under a given seed and configuration.

Failure paths are first-class. A missing observation, expired authorization,
queue overflow, unavailable service, tampered custody link or absent backup is
not a crash in the method; it is an expected branch with observable
consequences. In graph terms, each process may also be linked to the canonical
entity graph by responsibility and artifact edges:
\begin{equation}
E_P\subseteq \mathcal P\times R_P\times (V_0\cup V_r(t)),
\label{eq:process-edges}
\end{equation}
where $R_P$ contains process-view relationships such as \textsf{performed by},
\textsf{consumes}, \textsf{produces}, \textsf{reviews}, \textsf{blocks} and
\textsf{hands off}.

For the Car2X simulation contract, the process view can be instantiated as:
\begin{equation}
\begin{aligned}
p_1&=(\textsf{P-NOM},\textsf{nominal},\{veh_1,rsu_1,gw_1\},\\
&\quad\textsf{valid request},\textsf{request routed},\\
&\quad msg_r,msg_r,\varnothing),\\
p_2&=(\textsf{P-ATT},\textsf{attack},\{veh_1,rsu_1\},\\
&\quad\textsf{forged/replayed request},\textsf{request observed},\\
&\quad msg_r,msg_r,\varnothing),\\
p_3&=(\textsf{P-DET},\textsf{detection},\{gw_1\},\\
&\quad\textsf{request observed},\\
&\quad\textsf{trust result produced},msg_r,dec_t,\textsf{reject}),\\
p_4&=(\textsf{P-POL},\textsf{policy},\{gw_1,tc_1\},\\
&\quad\textsf{decision available},\\
&\quad\textsf{allow or block recorded},dec_t,dec_t,\textsf{block}),\\
p_5&=(\textsf{P-EVI},\textsf{evidence},\{gw_1\},\\
&\quad\textsf{decision recorded},\textsf{evidence preserved},\\
&\quad dec_t,ev_t,\varnothing).
\end{aligned}
\label{eq:car2x-processes}
\end{equation}
This example shows that the forged or replayed request is represented both as a
simulator event and as a process artifact. It is carried by an attack process,
reviewed by a detection process, constrained by a policy process and preserved
by an evidence process. The denial branch is explicit: the same process view can
show a blocked traffic-control influence without treating the simulation as a
failed run.

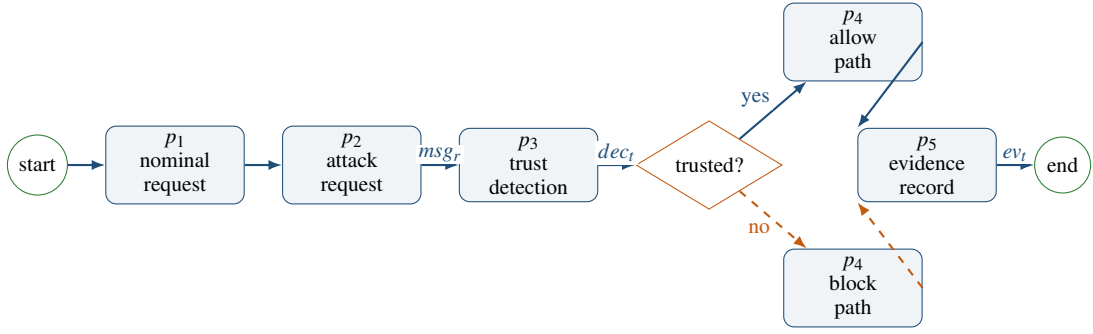
\begin{figure*}[t]
\centering
\begin{tikzpicture}[
  scale=.9,
  transform shape,
  node distance=9mm and 5.5mm,
  evt/.style={draw=salusgreen,circle,fill=white,minimum size=7mm,align=center,
  font=\small},
  task/.style={draw=salusblue,rounded corners,fill=salusgray,text width=18mm,
  minimum height=9mm,align=center,font=\small},
  gate/.style={draw=salusorange,diamond,aspect=1.65,fill=white,text width=16mm,
  inner sep=1pt,align=center,font=\small},
  flow/.style={-{Latex[length=1.8mm]},thick,salusblue},
  deny/.style={-{Latex[length=1.8mm]},thick,salusorange,dashed},
  elabel/.style={fill=white,inner sep=1pt,font=\small}]
\node[evt] (start) {start};
\node[task,right=of start] (nom) {$p_1$\\nominal\\request};
\node[task,right=of nom] (att) {$p_2$\\attack\\request};
\node[task,right=of att] (det) {$p_3$\\trust\\detection};
\node[gate,right=of det] (gate) {trusted?};
\node[task,above right=of gate] (polallow) {$p_4$\\allow\\path};
\node[task,below right=of gate] (polblock) {$p_4$\\block\\path};
\node[task,right=11mm of gate] (evi) {$p_5$\\evidence\\record};
\node[evt,right=of evi] (end) {end};
\draw[flow] (start)--(nom);
\draw[flow] (nom)--(att);
\draw[flow] (att)--node[above,elabel]{$msg_r$}(det);
\draw[flow] (det)--node[above,elabel]{$dec_t$}(gate);
\draw[flow] (gate)--node[above left,elabel]{yes}(polallow);
\draw[deny] (gate)--node[below left,elabel]{no}(polblock);
\draw[flow] (polallow.east)--(evi.north west);
\draw[deny] (polblock.east)--(evi.south west);
\draw[flow] (evi)--node[above,elabel]{$ev_t$}(end);
\end{tikzpicture}
\caption{BPMN-style Car2X process view matching the Section~\ref{sec:source-governed-case}
simulation contract. The dashed branch is the explicit denial path for a forged,
replayed or stale request.}
\label{fig:car2x-process-view}
\end{figure*}

\subsubsection{Process and Behaviour Semantics}

The process view becomes executable only after its steps are tied to event
semantics. This adds deterministic ordering, declared uncertainty, explicit
queues, observation loss and twin updates, while preserving the same visibility
boundary used in the process view:
\begin{equation}
H \longrightarrow O \longrightarrow T \longrightarrow E \longrightarrow C,
\qquad C\not\leftarrow H,
\label{eq:visibility}
\end{equation}
where $H$ is hidden truth, $O$ observable records, $T$ accepted twin state,
$E$ verified evidence and $C$ reconstructed claims. Hidden truth may evaluate a
synthetic run after the fact, but accepted twin states and reconstructed claims
are derived from visible observations and evidence only.

The shared event vocabulary is:
\begin{equation}
\begin{aligned}
\mathcal E&=\{\epsilon_k\mid\\
\epsilon_k&=(eid_k,t_k,\hat t_k,src_k,dst_k,\\
&\quad\kappa_k,cid_k,\nu_k,\chi_k)\}.
\end{aligned}
\label{eq:event-vocabulary}
\end{equation}
Here $eid_k$ is the event identifier, $t_k$ is the event time, $\hat t_k$ is the
ingest time, $src_k,dst_k\in V_0\cup V_r(t)$ are the source and destination,
$\kappa_k$ is the event type, $cid_k$ is the correlation identifier, $\nu_k$ is
the visibility class and $\chi_k$ is the schema or payload contract. Each
scenario must state which events exist, which entities produced or consumed
them, and whether records were visible, withheld, delayed or lost.

Accepted twin state is updated only from visible observations:
\begin{equation}
\begin{aligned}
T_{k+1}&=U(T_k,O_k;\Theta),\\
O_k&=\Omega(H_k,\epsilon_k;\Lambda).
\end{aligned}
\label{eq:twin-update}
\end{equation}
In this notation, $T_k$ is the accepted twin state at step $k$, $U$ is the twin
update rule, $O_k$ is the observation available to the twin, $\Theta$ contains
model parameters, $\Omega$ is the observation function and $\Lambda$ contains
loss, delay, retention and visibility settings.

Queues, losses and delays are explicit. For a queue with arrivals $A_k$, service
$S_k$, drops $D_k$ and occupancy $B_k$, the invariant is:
\begin{equation}
B_{k+1}=B_k+A_k-S_k-D_k,\quad
0\le B_{k+1}\le B_{max}.
\label{eq:queue}
\end{equation}
Here $A_k$ is the number of arriving events, $S_k$ is the number served,
$D_k$ is the number dropped, $B_k$ is queue occupancy and $B_{max}$ is capacity.
Recording why an event is missing matters because channel loss, observer loss,
queue overflow, retention expiry and policy denial have different security
meanings. In the Car2X example, the gateway twin therefore validates observable
message features, records the reason for an allow/block decision, and preserves
that reason as evidence without reading the simulator's private attack label.

\subsection{Security Governance}

The security governance layer covers detection, authorization, safety guards,
minimization, evidence, custody and reconstruction. Its role is to decide what
may be done with a detection result, which records may become evidence and which
claims may later be reconstructed. \framework{} treats alerts and
recommendations as inputs to governed decisions, not as automatic authority. A
detector may score an untrusted control message or an anomalous device-flow
pattern, but the applied decision is still constrained by capability,
authorization and safety:
\begin{equation}
\begin{aligned}
a_t^{req}&=\pi(alert_t,X_t,cap_t,auth_t),\\
a_t^{app}&=\Gamma(a_t^{req},cap_t,\\
&\quad auth_t,safe(X_t)).
\end{aligned}
\label{eq:policy}
\end{equation}
Here $alert_t$ is the detection output at time $t$, $X_t$ is the governance
state available to the policy, $cap_t$ records what the simulated system is
allowed to do, $auth_t$ records the applicable authorization and $safe(X_t)$ is
the safety predicate. The policy $\pi$ proposes a requested action
$a_t^{req}$, while the guard $\Gamma$ converts it into the applied action
$a_t^{app}$ after checking capability, authorization and safety.
The requested and applied actions are separate records. This separation allows
the framework to express scenario-specific guards: detecting an attack does not
automatically authorize the strongest available response.

In practice, $X_t$ is the information that the policy is allowed to see at the
moment of decision. It is not the simulator's hidden truth. It may include the
accepted twin state, the current validation result, whether the message is fresh,
whether the queue is overloaded, the current operating mode, the declared
simulation boundary and whether an evidence record can be preserved.

For Car2X, the policy receives a visible validation result for a roadside-unit
request. If the request is fresh, authorized and safe for the current controller
state, the policy may request that the simulated command be applied. If the
request is a replay, is stale, comes from an unauthorized source or asks for an
unsafe traffic-control effect, the requested action is to block it. The final
guard then checks the scenario boundary: because the experiment prohibits live
RF and real traffic-control actuation, the applied action is recorded as
\textsf{block-record}. The governance record keeps the request, validation
reason, allow/block decision and evidence record together, so a reviewer can see
both why the request was blocked and which visible evidence supports that
decision.

Evidence follows the same discipline. A source record becomes evidence only
after authorization, minimization, canonicalization and hashing. Custody links
record how evidence was handled, but the framework does not claim that a local
hash chain alone establishes legal admissibility. Reconstruction consumes
verified visible evidence, not hidden truth. A claim therefore carries its
supporting evidence, time interval, missing expected evidence, alternatives,
limitations and confidence or support score.

\subsection{Assurance Package}

The assurance package layer collects the experiment matrix, Monte Carlo design,
runtime graph, competency queries, tests, reports and manifest. These artifacts
make the framework falsifiable. A scenario package is not complete when the
simulator runs once; it is complete when its artifacts can be checked as a
chain. Table~\ref{tab:method} summarizes the common delivery method. The table
is intentionally procedural: it states what must exist before a result can be
reviewed, rerun or challenged.

\begin{table*}[t]
\centering
\scriptsize
\setlength{\tabcolsep}{3pt}
\caption{Common \framework{} delivery method for security-DT scenarios.}
\label{tab:method}
\begin{tabularx}{\textwidth}{L{13mm}L{38mm}Y}
\toprule
Stage & Gate & Required outcome \\
\midrule
1 & Source-governed case & Bounded scenario, source precedence,
assumptions, prohibited capabilities and integration boundary. \\
2 & Canonical model & Stable entities, relationships, zones, roles, twins,
data objects and schema-valid inventories. \\
3 & Process and behaviour & BPMN/lifecycle views, state variables, events,
queues, observation loss, twin updates and failure branches. \\
4 & Security governance & Detection, authorization, safety guards,
minimization, evidence, custody, reconstruction and policy metrics. \\
5 & Implementation & Deterministic simulator, named seeds, adapters, validation
rules and complete run export. \\
6 & Experiments & Nominal case, negative or degraded branches, expected
assertions and comparison report. \\
7 & Evidence graph & Runtime graph, evidence manifest, custody links,
reconstruction and competency-query answers. \\
8 & Uncertainty & Predeclared Monte Carlo design, null-aware summaries and
sensitivity interpretation. \\
9 & Verification & Schema, invariant, privacy, safety, determinism, tamper,
experiment and report-completeness tests. \\
10 & Handoff & Integration contracts, dependency lock, technical report,
validation matrix and SHA-256 artifact manifest. \\
\bottomrule
\end{tabularx}
\end{table*}

The important property is not identical domain logic across scenarios, but a
shared traceability spine. A reviewer should be able to start from a reported
result and find the requirement or assumption that
motivated it, the entities and process steps involved, the configuration and
seed that produced it, the evidence record that supports it, the tests that
checked it and the manifest entry that identifies the exact artifacts.

For the Car2X scenario, this means that a replayed-request experiment is not
reported only as ``blocked''. The assurance package also records which request
was observed, which validation reason was produced, why the policy applied
\textsf{block-record}, whether any queue or freshness condition limited the
result, whether the traffic controller remained unactuated and which evidence
object and manifest hash support the statement. If the message observation is
lost or delayed, the package should report an unsupported or limited claim
rather than presenting the run as a successful detection.

The final assurance-package step is claim discipline: limiting conclusions to
the evidence actually present in the package. A scenario may be
\status{planned}, \status{formalized},
\status{executable}, \status{internally validated}, \status{externally
calibrated} or \status{deployed}; these labels describe evidential maturity, not
the ambition of the use case.

In this paper, the implemented packages support internally validated synthetic
claims: the declared models, controls, evidence paths, experiments, tests and
reports behave as specified under stated synthetic assumptions. They do not
establish operational safety, legal compliance, standards conformance, radio
performance, industrial fidelity or pilot KPIs. For Car2X, for example, the
package may claim that a forged or replayed request was blocked within the
declared synthetic run and that visible evidence supports the decision. It may
not claim protection of a real road junction, nor successful detection when the
observation is missing or stale. Stronger claims require calibrated data,
confirmed authority, real integration boundaries and operational safety
evidence.

\section{Application to selected use cases}
\label{sec:applications}

A general assurance framework is useful only if it can be instantiated and
checked across scenarios with different assets, observations, risks and
authority constraints. The evaluation therefore asks whether the same
source-governed, evidence-bounded structure can be applied across heterogeneous
cyber-physical settings without erasing their domain-specific constraints. The
five use cases below were selected for this purpose because each stresses a
different part of the framework: trust before control, passive and
privacy-preserving observation, safety-constrained response, lawful
evidence acquisition, and continuity-aware mitigation.

The broad use-case titles are larger than any executable model can responsibly
claim. For evaluation, each one is reduced to a bounded scenario with
explicit assumptions and prohibited capabilities. Simulation is then needed to
exercise the framework rather than merely describe it: each scenario must
generate observations, decisions, evidence, degraded branches and claim limits
that can be checked against the assurance package. Table~\ref{tab:ucmap} fixes
the mapping used throughout the remainder of the paper and identifies the main
framework aspect tested by each scenario.

\begin{table*}[t]
\caption{\framework{} use-case numbers, bounded scenarios and the framework
properties exercised by each scenario.}
\label{tab:ucmap}
\centering
\small
\begin{tabularx}{\textwidth}{L{10mm}L{37mm}L{58mm}Y}
\toprule
UC & Name & Modelled scenario & Framework aspect tested \\
\midrule
UC1 & Identification and penetration of attack surfaces during Car2X
communication & A forged or replayed abstract request crossing an OBU--RSU trust
boundary toward a safety-gated traffic-signal controller. & Trust validation,
authority and fail-safe separation from control effects. \\
UC2 & On-the-spot detection and categorization of IoT devices in public spaces
& Passive, pseudonymized multi-interface observation, association,
classification and approval checking for a synthetic unapproved device. &
Observation ambiguity, privacy limits and hypothesis-level twins. \\
UC3 & Industrial cyberattacks and safety anomalies in nuclear/radiation emergency
contexts & A non-sensitive one-vessel control surrogate subjected to symbolic
protocol-aware fuzzing, unauthorized writes and bounded application load. &
Hybrid process dynamics, safety guards and symbolic industrial attacks. \\
UC4 & Lawful forensic investigation in a smart home for IoT-enabled crime &
Scoped CPE/Home Assistant evidence acquisition after unauthorized entry and
door-sensor tampering, with authorization, minimization and custody. & Full
evidence chain, authorization, minimization, custody and reconstruction. \\
UC5 & Monitoring and safeguarding connected medical devices in a hospital & A
synthetic patient-monitor flow anomaly whose response is constrained by clinical
continuity, followed by redacted authorized evidence sharing. & Safety-critical
mitigation, continuity constraints and redacted evidence sharing. \\
\bottomrule
\end{tabularx}
\end{table*}

The subsections below state how each use case instantiates the framework. They
do not yet report performance results; instead, they identify the domain
specialization, canonical entities, executable implementation and expert-owned
parameters needed before the simulations in Section~\ref{sec:results} can be
interpreted. To make the implementation logic inspectable without overloading
the main text, UC1 is used as the detailed worked example: its entities,
relationships, deterministic configuration, experiment branches and validation
evidence are summarized in Section~\ref{sec:applications} and reported in the
Appendix. For space reasons, UC2--UC5 are presented in a
shorter but comparable form, focusing on the scenario description, evaluation
role, model content and expert parameters. The complete executable packages,
proof artifacts, test evidence, configurations and generated reports for all
use cases are available through the project repository: \projectreporef.

\begin{table*}[!t]
\centering
\scriptsize
\setlength{\tabcolsep}{2pt}
\caption{UC1 canonical entities used by the executable package.}
\label{tab:uc1-entities}
\begin{tabular}{L{18mm}L{31mm}L{27mm}L{37mm}L{13mm}L{24mm}}
\toprule
ID & Name & Class & Role in implementation & Status & Assumption(s) \\
\midrule
UC1 & Car2X scenario & UseCase & Bounds the controlled scenario & proposal & UC1-001 \\
UC1-ENV01 & Controlled environment & Physical environment & Contains synthetic vehicle and roadside roles & proposal & UC1-002 \\
UC1-NZ01 & In-vehicle network & NetworkZone & Separates OBU/gateway from vehicle controllers & assumed & UC1-003 \\
UC1-NZ02 & Car2X channel zone & NetworkZone & Abstract V2I communication boundary & proposal & UC1-004 \\
UC1-NZ03 & Roadside backhaul & NetworkZone & Connects RSU and traffic-control services & assumed & UC1-005 \\
UC1-NZ04 & Security services zone & Network zone & Hosts detection, policy, twin and evidence services & eng. & UC1-006 \\
UC1-D01 & Police test vehicle & ConnectedVehicle & Protected demonstration platform & proposal & UC1-007 \\
UC1-D02 & On-board unit & Vehicle comms unit & Vehicle Car2X endpoint & proposal & UC1-004; UC1-008 \\
UC1-D03 & Vehicle gateway & Gateway & Abstract boundary to in-vehicle systems & assumed & UC1-003 \\
UC1-D04 & In-vehicle controller & Controller & Bounded downstream cyber target & assumed & UC1-009 \\
UC1-D05 & Roadside unit & RoadsideUnit & Infrastructure endpoint and validation point & proposal & UC1-010 \\
UC1-D06 & Traffic-signal controller & Controller & Evaluates accepted requests under safety interlocks & assumed & UC1-011 \\
UC1-D07 & Traffic signal & Actuator & Displays simulated controller phase & proposal & UC1-011 \\
UC1-D08 & Adversarial transmitter & Test asset & Injects symbolic forged, replayed or flood messages & eng. & UC1-001 \\
UC1-D09 & Observation component & Monitoring asset & Produces bounded message and flow observations & eng. & UC1-012 \\
UC1-S01 & Traffic-management service & Service & Provides supervisory infrastructure context & assumed & UC1-005 \\
UC1-S02 & PKI/trust service & Security service & Produces abstract credential, freshness and authorization decisions & proposal & UC1-013 \\
UC1-S03 & Detection service & Security service & Scores validation and traffic features & eng. & UC1-014 \\
UC1-S04 & Policy service & Security service & Requests mitigation and records acknowledgement & proposal & UC1-015 \\
UC1-S05 & Twin synchronization & DT service & Maintains versioned observable entity state & proposal & UC1-016 \\
UC1-S06 & Evidence/custody service & Forensics service & Minimizes, hashes and registers evidence & eng. & UC1-017 \\
UC1-S07 & Reconstruction service & Forensics service & Builds evidence-linked claims & eng. & UC1-018 \\
UC1-A01 & Vehicle operator & HumanActor & Operates the synthetic test drive & proposal & UC1-019 \\
UC1-A02 & Infrastructure operator & Organisation actor & Owns or supervises roadside assets & proposal & UC1-020 \\
UC1-A03 & Red-team operator & Human actor & Initiates the controlled symbolic attack & eng. & UC1-001 \\
UC1-A04 & LEA analyst & LEAUser & Reviews minimized evidence and reconstruction & proposal & UC1-017 \\
UC1-DT01 & Vehicle DT & TwinEntity & Mirrors observable vehicle state & proposal & UC1-016 \\
UC1-DT02 & OBU DT & TwinEntity & Mirrors observable OBU state & proposal & UC1-016 \\
UC1-DT05 & RSU DT & TwinEntity & Mirrors observable RSU state & proposal & UC1-016 \\
UC1-DT06 & Signal-controller DT & TwinEntity & Mirrors observable controller state & proposal & UC1-016 \\
\bottomrule
\end{tabular}
\end{table*}

\begin{table*}[!t]
\centering
\scriptsize
\setlength{\tabcolsep}{2pt}
\caption{UC1 functional relationships, excluding simple belongs-to-use-case
membership edges.}
\label{tab:uc1-relationships}
\begin{tabular}{L{20mm}L{24mm}L{24mm}L{25mm}L{54mm}}
\toprule
ID & Source & Relation & Target & Implementation meaning \\
\midrule
UC1-REL-006 & UC1-D01 & DEPLOYED\_IN & UC1-ENV01 & Vehicle is placed in the controlled environment. \\
UC1-REL-007 & UC1-D05 & DEPLOYED\_IN & UC1-ENV01 & RSU is placed in the controlled environment. \\
UC1-REL-008 & UC1-D02 & communicates with & UC1-NZ02 & OBU uses the abstract Car2X channel. \\
UC1-REL-009 & UC1-D05 & communicates with & UC1-NZ02 & RSU receives through the abstract Car2X channel. \\
UC1-REL-010 & UC1-D02 & communicates with & UC1-D05 & OBU and RSU exchange abstract requests. \\
UC1-REL-011 & UC1-D05 & validated by & UC1-S02 & RSU messages use trust validation. \\
UC1-REL-012 & UC1-D05 & routes to & UC1-D06 & Accepted requests may reach the controller. \\
UC1-REL-013 & UC1-D06 & CONTROLS & UC1-D07 & Controller determines simulated signal phase. \\
UC1-REL-014 & UC1-D08 & TARGETS & UC1-D05 & Symbolic transmitter targets the RSU. \\
UC1-REL-015 & UC1-D08 & attempts to trigger & UC1-D06 & Attack attempts a controller effect. \\
UC1-REL-016 & UC1-D09 & FEEDS & UC1-S03 & Observations feed detection. \\
UC1-REL-017 & UC1-S02 & FEEDS & UC1-S03 & Trust results feed detection. \\
UC1-REL-018 & UC1-S03 & TRIGGERS & UC1-S04 & Alerts trigger policy decisions. \\
UC1-REL-019 & UC1-S04 & mitigated by & UC1-D05 & Mitigation is applied at the RSU boundary. \\
UC1-REL-020 & UC1-D06 & protected by & UC1-S04 & Controller path is policy-protected. \\
UC1-REL-021 & UC1-D01 & mirrored by & UC1-DT01 & Vehicle has an observable DT. \\
UC1-REL-022 & UC1-D02 & mirrored by & UC1-DT02 & OBU has an observable DT. \\
UC1-REL-023 & UC1-D05 & mirrored by & UC1-DT05 & RSU has an observable DT. \\
UC1-REL-024 & UC1-D06 & mirrored by & UC1-DT06 & Controller has an observable DT. \\
UC1-REL-025 & UC1-S05 & synchronizes with & UC1-DT01 & Twin service updates vehicle twin. \\
UC1-REL-026 & UC1-S05 & synchronizes with & UC1-DT02 & Twin service updates OBU twin. \\
UC1-REL-027 & UC1-S05 & synchronizes with & UC1-DT05 & Twin service updates RSU twin. \\
UC1-REL-028 & UC1-S05 & synchronizes with & UC1-DT06 & Twin service updates controller twin. \\
UC1-REL-029 & UC1-S03 & produces evidence & UC1-S06 & Alerts can become evidence. \\
UC1-REL-030 & UC1-S02 & produces evidence & UC1-S06 & Trust decisions can become evidence. \\
UC1-REL-031 & UC1-S04 & produces evidence & UC1-S06 & Policy decisions can become evidence. \\
UC1-REL-032 & UC1-S06 & FEEDS & UC1-S07 & Verified evidence feeds reconstruction. \\
UC1-REL-033 & UC1-S07 & FEEDS & UC1-A04 & Reconstruction is reviewed by analyst. \\
UC1-REL-034 & UC1-A01 & OPERATED\_BY & UC1-D01 & Authorized operator runs the vehicle. \\
UC1-REL-035 & UC1-A02 & OWNED\_BY & UC1-D05 & Infrastructure operator owns the RSU role. \\
UC1-REL-036 & UC1-A03 & OPERATED\_BY & UC1-D08 & Red-team operator controls the symbolic transmitter. \\
\bottomrule
\end{tabular}
\end{table*}

\begin{table*}[!t]
\caption{UC1 deterministic scenario configuration used for the nominal
forged-request run.}
\label{tab:uc1-config}
\centering
\small
\begin{tabularx}{\linewidth}{L{34mm}L{78mm}Y}
\toprule
Configuration area & Implemented value & Purpose in the UC1 evaluation \\
\midrule
Scenario and seed & UC1 forged-control affecting-request; seed 1101; 120~s & Makes the run deterministic and reproducible. \\
Attack schedule & Forged request; start 45~s; count 3; interval 0.1~s; untrusted credential & Exercises trust validation and policy before control. \\
Channel & 20 messages/s; buffer 20; 20~ms base delay; 5~ms jitter; zero nominal loss; observation probability 1.0 & Makes queueing, delay and observation explicit. \\
Trust validation & Service available; freshness window 2~s; valid sender \texttt{UC1-D02} & Separates credential, freshness and sender authority. \\
Detector & Threshold 0.4; weights auth 0.45, freshness 0.25, and sequence, rate and plausibility 0.10 each & Produces explainable alerts rather than opaque scores. \\
Policy & Supported actions drop, rate-limit, isolate, alert-only; selected action drop; enforcement delay 0.05~s & Tests requested versus applied mitigation. \\
Controller safety & Initial phase safe-all-red; minimum phase 3~s; control requests synthetically allowed & Ensures no controller effect occurs without accepted authority. \\
Evidence & Enabled; message, trust, alert, policy and controller records; 3600~s retention & Defines what can support claims and reconstruction. \\
Clock offsets & OBU 0.01~s, RSU -0.01~s, detector 0.0~s & Makes time provenance explicit. \\
\bottomrule
\end{tabularx}
\end{table*}

\begin{table*}[!t]
\caption{UC1 experiment matrix used to test nominal, degraded and negative
branches.}
\label{tab:uc1-experiments}
\centering
\small
\begin{tabularx}{\linewidth}{L{34mm}L{53mm}Y}
\toprule
Experiment branch & Perturbation & Expected framework property \\
\midrule
baseline & Attack disabled & No alert or safety violation is invented. \\
nominal forged request & Base configuration & Alerts occur; controller effects remain zero. \\
expired certificate & Credential state expired & Trust failure is detected and blocked. \\
revoked certificate & Credential state revoked & Revocation is treated as a trust failure. \\
replay outside window & Replay, valid credential, replay age 10~s & Freshness violation prevents authority. \\
message flood & 50 messages, 0.01~s interval, buffer 3, service 5 messages/s & Queue loss is explicit; no safety violation. \\
trust unavailable & Trust service unavailable & Dependency failure fails closed. \\
policy enforcement failed & Enforcement unavailable & Alert and safety records remain inspectable. \\
observation loss & Observation probability 0.0 & Hidden truth does not create alerts or claims. \\
\bottomrule
\end{tabularx}
\end{table*}

\begin{table*}[!t]
\caption{UC1 validation evidence showing why the implementation can support an
internal framework-validity claim.}
\label{tab:uc1-validation}
\centering
\small
\begin{tabularx}{\linewidth}{L{38mm}L{24mm}Y}
\toprule
Requirement area & Status & Evidence retained in the package \\
\midrule
Scenario, sources and assumptions & Complete & Use-case document and assumption register; partner confirmation remains calibration work. \\
Entities and stable IDs & Complete & \texttt{models/instances/uc1\_entities.json}; schema tests. \\
Relationships and graph & Complete & 60 relationships; graph tests; \texttt{models/graph/uc1\_graph.json}. \\
Formal model and BPMN views & Complete & Formal model plus seven BPMN process files. \\
Deterministic simulation & Complete & Simulation, reproducibility tests and generated run artifacts. \\
Trust, detection and policy & Complete & Validation events, explanation records and negative branches. \\
Controller safety & Complete & Zero violations across tests and experiments; not a safety certification. \\
Evidence, custody and reconstruction & Complete & Hashing, minimization, tamper tests and evidence-linked reconstruction. \\
Experiment and uncertainty matrix & Complete & Nine passing branches and 100 predeclared synthetic samples. \\
Clean-environment validation & Partial & Lock file and passing tests; independent CI/platform run remains future work. \\
\bottomrule
\end{tabularx}
\end{table*}

\subsection{UC1: Car2X attack-surface modelling}

\textbf{Scenario description.} \textit{A police test vehicle approaches a signalized junction and uses its
on-board Car2X unit to exchange priority-related messages with roadside
infrastructure. During the same run, a red-team transmitter injects forged or
replayed messages that appear structurally plausible but should not be allowed
to influence the traffic-signal controller. The digital twin must therefore
follow the message, decide what is observable, validate trust and freshness,
record any security decision, and show that an untrusted request was contained
before it became a physical control effect.}

\textbf{Evaluation role.} UC1 tests whether the framework can distinguish an
observed communication attack from authority to influence a physical controller.
It is the trust-and-control case: the important outcome is not only that an
attack is detected, but that the framework can explain why a visible request is
not equivalent to an authorized command. The scenario therefore checks the full
chain from observed message to validation result, policy action, controller
guard and evidence-backed reconstruction.

\textbf{Definition.} The simulated scenario contains a synthetic police
test vehicle, its on-board unit (OBU), a roadside unit (RSU), a bounded
traffic-signal controller and supporting trust, detection, policy, twin and
evidence services. The OBU and RSU exchange abstract authenticated requests
over a declared Car2X channel; the controller is reachable only through the RSU
and only after validation and policy checks. The adversarial transmitter does
not compromise a real vehicle or radio stack. It injects symbolic forged,
replayed or flooded messages into the same abstract channel so that the
framework can test denial, degradation and evidence paths under controlled
conditions. No live radio, vehicle or traffic-control interface exists.

\textbf{Entities.} The canonical model contains 30 entities and 60 typed
relationships: vehicle, OBU, gateway, RSU, in-vehicle and signal controllers,
traffic signal, adversarial transmitter, road-and-lab environment, four zones,
operator, red-team and analyst roles, trust, detection, policy, twin and
evidence services,
and four physical--digital pairs. These entities make the scenario auditable:
each message has a producer and receiver, each validation or policy decision is
attached to the service that made it, each possible controller effect is linked
to a protected physical endpoint, and each claim can be traced back to evidence
rather than hidden simulator truth. The implementation-level entity inventory,
functional relationships, deterministic configuration, experiment branches and
validation evidence are reported in the Appendix.

\textbf{DT implementation.} The execution is organized as a
trust-gated message pipeline. Messages first enter a finite channel queue, so
delay, congestion and loss are explicit rather than implicit. Second, the RSU
and trust service evaluate whether the message can be parsed, whether the trust
service is available, whether the credential is valid, whether the sender is
authorized, and whether the timestamp, sequence and replay window are
acceptable. The detector scores the visible validation and traffic
features with declared weights, producing an explainable alert rather than an
opaque label. The policy service translates the alert into an allowed
mitigation such as drop, rate-limit, isolate or alert-only, and records whether
that mitigation was actually applied. Finally, the controller safety gate accepts
only messages that have survived the trust and policy path; rejected or
untrusted requests can still be recorded as attempted transitions, but they do
not change the controller state.

The DT state follows the same visibility discipline as the general framework.
The vehicle, OBU, RSU and signal-controller twins are updated only from observed
message and state events. The simulator may know that an injected message is an
attack, but the operational twin and reconstruction logic cannot read that
hidden label directly. They must infer from visible credentials, freshness,
sequence behaviour, validation results, alerts and policy records. Evidence is
then minimized, hashed and custody-linked, allowing the final claim to state, for
example, that a forged request was observed, rejected and prevented from
producing a controller effect within the declared synthetic model.

\textbf{Expert parameters.} UC experts must confirm the Car2X profile
(ITS-G5/C-V2X), message taxonomy and permissions, PKI semantics, radio/channel
calibration, OBU/RSU/controller topology and telemetry, controller phases and
minimum times, enforcement capabilities, evidence authority/retention and the
role of the DS1 in-vehicle dataset. These replace assumptions through versioned
configuration or adapters without weakening safety and truth separation.

\subsection{UC2: Passive detection and categorization of IoT devices}

\textbf{Scenario description.} \textit{A public-space-inspired synthetic
monitoring scene represents the kind of setting targeted by UC2, such as a
transport hub, campus area or event venue, without modelling a real location.
Three passive observers generate abstract 3GPP, Wi-Fi and Bluetooth
observations from approved device classes and from one unapproved device
hypothesis. The system must support situational awareness in this public-space
domain, but it must not turn uncertain radio observations into confirmed
identity, capture payloads or trigger active interference.}

\textbf{Evaluation role.} UC2 tests whether the framework can model partial
observation without converting uncertain passive signals into confirmed
identity. It exercises hypothesis twins, privacy-preserving evidence and
fail-closed behaviour under loss, ambiguity and stale registry state.

\textbf{Definition.} The model represents a synthetic public-space observation
area rather than a private home, vehicle, industrial plant or real venue.
Three approved device classes and one unapproved device hypothesis emit a small
deterministic set of abstract 3GPP, Wi-Fi or Bluetooth observations. Passive
observers pseudonymize metadata; analytics associates observations, estimates a
device class, consults an approval registry and requests bounded review. Active
interception, payload capture, subscriber identity and location tracking are
prohibited.

\textbf{Entities.} The canonical model contains 25 entities and 26
relationships: three passive observers and their twins, approved and unapproved
device classes and hypothesis twins, scene, analytics and forensic zones,
association and classification components, registry, explainable detection,
policy, consultation and forensic services, stakeholder and human roles, and
synthetic and future-data boundaries.

\textbf{DT implementation.} The executable model treats each
passive radio observation as a lossy and uncertain record. Whether a device is
seen depends on the declared coverage of the observer, the configured loss rate
and the interface being monitored. The observed features are intentionally
noisy: they support association and classification, but they do not expose
payloads, subscriber identity or precise location. Observations can be
associated only when they fall within the declared time window and interface
constraints. The classifier then returns a most likely device class, alternative
classes and a confidence value; the approval state is represented separately as
approved, unapproved, unknown or stale. Here the twin referent is explicitly a
device hypothesis as well as observer state, so hypothesis fidelity and
ambiguity are reported. The deterministic engine separates restricted emitter
truth from lossy passive observations, enforces a bounded queue, exports class
alternatives and confidence, and permits only non-RF review responses. Evidence
is minimized, hashed and custody-linked; reconstruction uses evidence rather
than hidden truth.

\textbf{Expert parameters.} Experts must supply permitted observation fields,
legal/ethical authority, event topology, interface generations/bands, device
taxonomy, approval-list governance, coverage/loss/delay and clock profiles,
pseudonym rotation, association window, classifier calibration/confusion,
thresholds, response capabilities, retention and DS2/DS3 licensing/mapping.

\subsection{UC3: Industrial cyber-physical surrogate}

\textbf{Scenario description.} \textit{A synthetic industrial control cell
regulates the level of a single process vessel using a level sensor, PLC, inlet
pump and outlet valve. The HMI, engineering workstation, historian, industrial
gateway and passive monitor are represented as supervisory and control-network
assets around that process. During the run, configured symbolic events exercise
smart fuzzing, unauthorized write attempts, application load, sensor-fault and
observation-loss branches. The twin must follow both the cyber events and the
evolving process state, but any mitigation must respect safe operating limits
and scan-cycle control rather than react to detection alone.}

\textbf{Evaluation role.} UC3 tests the framework on a hybrid process-control
setting in which cyber events may have physical consequences. Its value is to
exercise process state, scan-cycle control, symbolic protocol checks and
safety-constrained mitigation without claiming nuclear-plant fidelity.

\textbf{Definition.} A deterministic one-vessel level-control surrogate models
a sensor, PLC, pump, outlet valve, HMI, engineering workstation, historian,
gateway and passive monitoring. Attack messages are scheduled from
configuration and validated without constructing real protocol payloads or
opening industrial-control adapters. Symbolic protocol-aware fuzzing,
unauthorized writes, bounded application load and sensor-fault branches are
tested; any mitigation is constrained by process safety. The model does not
expose or represent an operational nuclear facility.

\textbf{Entities.} There are 36 canonical entities and 60 relationships,
including nine physical/logical twins, four zones, process and control assets,
security and forensic services, stakeholder roles and synthetic/SWaT/operations
data boundaries.

\textbf{DT implementation.} The executable model updates the vessel level from
configured inflow, outflow and actuator settings while keeping the value within
declared bounds. The PLC reads the level at scan boundaries and applies a simple
hysteretic control rule, so commands do not change continuously or outside the
control cycle. Cyber requests traverse a bounded queue and symbolic
protocol/authorization checks before they can influence detector scores or
policy decisions. The nine twins accept only versioned observations, and the
reconstruction step is built from retained evidence rather than from hidden
simulator truth.

\textbf{Expert parameters.} Experts must confirm a non-sensitive abstract
topology, vessel, flow, sensor and scan parameters, safe envelopes and interlocks,
protocol and request abstraction, observation coverage, gateway capability,
detection/policy calibration, retention and human responsibilities. SWaT and
operations datasets remain unavailable until licensing, checksum, field mapping
and scope review are complete.

\subsection{UC4: Smart-home lawful forensic acquisition}

\textbf{Scenario description.} \textit{A synthetic smart home records an
unauthorized door opening followed shortly by door-sensor tampering and
disconnection. Home Assistant may trigger an entry-light automation, while CPE
flow records and device state events provide partially overlapping source
evidence. A case trigger starts the investigation workflow, but evidence may be
collected only through scoped authorization, minimization, custody and
reconstruction rules.}

\textbf{Evaluation role.} UC4 is the evidence-chain stress test. It exercises
the largest graph, the deepest evidence model and the strongest reconstruction
requirements, including explicit authorization, minimization, custody, tamper
handling and qualified claims under missing or degraded sources.

\textbf{Definition.} Unauthorized door entry and door-sensor tampering occur in
a synthetic armed, unoccupied smart home. Home Assistant may trigger a light;
CPE and application records overlap imperfectly. A case trigger remains separate
from authorization. A valid scoped request collects minimized records through
an abstract lawful-interception module, preserves custody and supports a
qualified reconstruction.

\textbf{Entities.} The model contains 38 entities and 123 relationships:
premises, CPE, Home Assistant, door and presence sensors, smart plug, smart
light, optional cloud, six twins, resident, unauthorized, investigator, auditor
and authority roles, policy, alert and data records, evidence, custody and
reconstruction services across three zones. Nine common schemas are augmented
by nine legacy device inventory schemas.

\textbf{DT implementation.} UC4 instantiates the full evidence workflow. The
simulation keeps imperfect sensing, source clocks, CPE flow consistency and
monotonic twin versions visible. Authorization is checked before acquisition
using the declared purpose, validity, evidence window, entities, record types
and requested action. Before disclosure, records are projected to the permitted
fields, hashed in canonical form, linked into custody and used to build an
interval-based reconstruction with explicit alternatives.

\textbf{Expert parameters.} Experts must confirm device makes and capabilities,
protocols, CPE and Home Assistant APIs and retention, evidentiary fields, real
authorization and purpose semantics, network, clock and sensor calibration, cloud
scope, identity handling and calibrated claim weights. Current legal and privacy
controls are executable safeguards, not a legal opinion or admissibility claim.

\subsection{UC5: Clinical-continuity-aware medical IoT monitoring}

\textbf{Scenario description.} \textit{A synthetic hospital patient monitor
emits normal abstract flows and then shows anomalous egress to a new
destination. In the nominal run the monitor is essential and no approved backup
monitoring is active, so an isolation request must be rejected even when an
alert is raised. A separate backup-active branch shows when bounded isolation
becomes eligible. The framework must show whether a security response is
permitted only when clinical continuity and evidence-sharing constraints are
satisfied.}

\textbf{Evaluation role.} UC5 tests whether governance can encode a
domain-specific safety priority that overrides a plausible security response.
It exercises requested-versus-applied mitigation, continuity constraints,
redaction and authorized sharing in a clinical setting.

\textbf{Definition.} A synthetic patient monitor produces normal abstract
flows. A symbolic compromise produces anomalous egress and inconsistent health
metadata. Detection and policy must preserve essential monitoring: isolation is
ineligible in the no-backup nominal configuration and eligible only in the
backup-active branch. Evidence is minimized, redacted and shared only under
explicit authority. No patient data, diagnosis, treatment or medical actuation
is permitted.

\textbf{Entities.} The canonical model contains 34 entities and 45
relationships: a patient monitor and inactive future device classes,
gateway, switch, collector and console assets, four zones, registry,
explainable detection, clinical-safety policy, evidence and custody,
redaction and sharing, reconstruction and threat-context services, seven twins,
synthetic and future-data boundaries, and clinical, security and investigator
stakeholder roles.

\textbf{DT implementation.} The executable DT records observable device health,
flow behaviour, destination changes, care dependency, freshness and safety
decisions. Isolation is allowed only when the request is authorized, the
capability exists and continuity of essential monitoring is preserved through an
approved backup or a non-essential device status. The implementation separates
restricted attack truth from observations, exports feature contributions and
alternatives, records requested versus applied response, rejects unsafe
isolation, verifies evidence and custody, redacts shared packages and
reconstructs from evidence only.

\textbf{Expert parameters.} Hospital and UC experts must define device and zone
inventory, vendor-neutral flow/health contracts, clinical criticality and backup
criteria, normal-flow distributions, attack abstraction, observation and clock
profiles, detection calibration, mitigation rights and review, evidence purpose,
redaction/sharing authority, retention, stakeholder ownership, and DS8 licensing
and field mapping.

\section{Simulation Design and Results}
\label{sec:results}

The use-case descriptions above define what each scenario must represent, which
entities and relationships are in scope, and which expert assumptions still
bound the model. The next step is to test whether these descriptions can be
executed as consistent DT packages rather than remaining conceptual examples.
For this reason, the evaluation focuses on simulation evidence: each scenario is
run under nominal, degraded and negative conditions, and the resulting artifacts
are checked against the framework's requirements for observation boundaries,
policy decisions, evidence handling and claim discipline.

The results should therefore be read as framework-validity evidence. They show
whether the proposed modelling chain can be instantiated across different
application domains, whether the generated simulations expose the intended
guards and failure modes, and whether the final claims are supported by
observable records rather than hidden simulator truth. They do not claim field
performance, operational calibration, legal admissibility, clinical safety or
industrial fidelity.

\subsection{Evaluation Design and Evidence Status}

Each use case follows the same experimental structure. A baseline or nominal
scenario checks the expected path through the model. Negative or degraded
branches then test whether missing observations, unavailable services, invalid
authority, queue pressure, tampering or unsafe response conditions are
represented as explicit outcomes. A 100-sample uncertainty design varies
declared synthetic parameters to show which assumptions dominate the reported
metrics. Finally, schema, invariant, determinism, privacy/safety and report
completeness tests check whether the generated artifacts support the stated
claim.

All simulations are deterministic and synthetic. Their experiment matrices
probe designed edge cases, while the 100-run analyses vary declared uncertain
parameters. Results are descriptive model outcomes and are not statistical
estimates of field performance. ``Complete'' therefore means complete against
the internal implementation contract, not externally calibrated or deployed.

Table~\ref{tab:status} summarizes the implementation evidence available for
each use case. Its purpose is to show that all scenarios were implemented and
internally validated against the same artifact contract, while still allowing
different numbers of entities, relationships, schemas, process views and tests
according to domain complexity.

\begin{table}[ht]
\caption{Validated implementation summary for the five completed use cases.}
\label{tab:status}
\centering
\small
\begin{tabular}{lrrrrr}
\toprule
UC & Entities & Relations & Core schemas & BPMN & Tests \\
\midrule
UC1 & 30 & 60 & 9 & 7 & 22 \\
UC2 & 25 & 26 & 9 & 7 & 11 \\
UC3 & 36 & 60 & 9 & 8 & 21 \\
UC4 & 38 & 123 & 9$+9$ & 7 & 47 \\
UC5 & 34 & 45 & 9 & 8 & 13 \\
\bottomrule
\end{tabular}
\end{table}

Together, the implementations comprise 163 canonical entities, 314 typed
relationships, 37 BPMN artefacts, 52 passing experiment branches, five
100-sample uncertainty analyses and 114 passing tests. Each package also
contains source and configuration documentation, inventories and data mappings,
restricted-data boundary instructions, schema-valid event and evidence fixtures,
integration contracts, a compiled technical report, retained validation
evidence, a final handoff and a verified SHA-256 manifest.

\subsection{Simulation and Results}

The following subsubsections report the simulation design, observed results and
framework-validity interpretation for each use case. The ordering follows the
use-case descriptions in Section~\ref{sec:applications}, so that each result can
be read against the corresponding scenario assumptions and model boundaries.

\subsubsection{UC1: Trust Before Control}

\textbf{Simulation design.} UC1 uses three nominal requests and an abstract
channel. It applies fail-closed trust validation, transparent weighted
detection, bounded policy actions and a signal safety interlock. Nine branches
cover baseline, forged requests, expired or revoked credentials, replay, flood,
trust-service unavailability, enforcement failure and complete observation loss.

\textbf{Results.} In the nominal run, all 3 attack messages were observed and
alerted (recall 1.0); validation and policy rejected them; controller effects,
safety violations, integrity failures and unauthorized disclosures were all
zero; evidence capture recall was 1.0. In the flood branch, 50 attacks were
generated, only 3 observed and 47 dropped by queue overflow, producing recall
0.06. Under complete observation loss, no alert was emitted. Across 100
predeclared samples, mean attack-detection recall was 0.28 (normal-approximation
95\% interval 0.200--0.360), evidence capture recall was 1.0, and controller
effects and safety violations remained zero.

\textbf{Framework-validity interpretation.} The implementation demonstrates fail-safe separation of
untrusted requests from controller effects. It also demonstrates that perfect
nominal recall depends on ideal observation and queue conditions: when coverage
and capacity are varied, detection drops while the safety gate still prevents
controller effects.

\subsubsection{UC2: Passive Observation and Privacy}

\textbf{Simulation design.} The versioned configurations define a 60-s passive scene
with three approved devices, one nominal unapproved device, four emissions per
device, three interface families, 0.95 coverage, 0.02 loss, a 2-s association
window, capacity 50, classifier accuracy parameter 0.90 and threshold 0.70. The
nine configured branches cover baseline, nominal, single-interface, observation
loss, overflow, registry unavailable/stale, low confidence and policy
unavailability. A 100-sample design varies coverage, loss, association window,
classifier accuracy and threshold.

\textbf{Results.} The nominal run delivered 45 observations, recorded three
lost observations, formed four classifications and raised one explainable alert.
Queue overflow, process effects, safety violations, active RF actions and
unauthorized disclosures were zero. All nine branches passed their assertions:
complete observation loss and severe queue overflow suppressed alerts rather
than inventing certainty; unavailable/stale registry and low-confidence cases
also failed closed. Across 100 predeclared samples, mean alert count was 0.60;
active RF actions, unauthorized disclosures and safety violations remained zero.
Eleven retained tests passed.

\textbf{Framework-validity interpretation.} UC2 demonstrates the framework's passive and privacy
invariants and makes information loss visible. Its alert behavior is an
internally consistent synthetic baseline, not evidence of field identification
accuracy or lawful operational authority.

\subsubsection{UC3: Safety-Constrained Industrial Response}

\textbf{Simulation design.} UC3 selected a deliberately non-sensitive one-vessel
surrogate with fixed scan-boundary control. Six symbolic attack requests drive
the nominal scenario. Nine branches distinguish unauthorized write, application
flood, sensor fault, observation loss, enforcement failure, unsafe isolation and
clock skew from baseline. Safety is enforced independently of alerting.

\textbf{Results.} The nominal run generated and validated 6 attack messages and
6 alerts, with attack-detection recall 1.0. Process level remained at 5.0~m in
the retained nominal output; process effects, safety violations, integrity
failures and unauthorized disclosures were zero. All nine branch assertions
passed. Application flood produced four alerts; complete observation loss
produced none. Across 100 runs, mean alert count was 3.88, while mean process
effects and total safety violations were zero. Twenty-one retained tests passed.

\textbf{Framework-validity interpretation.} The model verifies its architectural safety invariant and
attack/fault branch mechanics, not industrial-plant realism. A constant nominal
level and zero effects show containment in this abstraction; they do not
validate process fidelity or real detection quality.

\subsubsection{UC4: Lawful Evidence and Reconstruction}

\textbf{Simulation design.} UC4 fixed a 1,800-s horizon: warm-up at 300~s, entry at
600~s, tampering and disconnection at 610~s, case trigger at 900~s and acquisition at
920~s. Case and authority are separate. Payload capture is disabled; every
record is independently checked and minimized. Sixteen branches probe
authorization absent/invalid/expired/revoked, out-of-scope data, missing events,
record eviction, CPE overflow/unavailability, clock skew, evidence/custody
tampering and Home Assistant failure.

\textbf{Results.} The nominal branch collected and verified 17/17 relevant
items (capture recall and evidence precision 1.0), released zero unauthorized
bytes, and produced three strong claims. Collection p95 was 0.016~s and
investigation latency 20.017~s. Reconstruction recall was 1.0, precision 0.5,
timestamp MAE 0.0424~s, mean twin lag 0.488~s, acquisition utilization 0.190 and
peak acquisition occupancy 4,040 bytes. Precision 0.5 reflects ten reconstructed
relevant candidates for five matched truth events, and must not be hidden by the
perfect recall.

Negative branches behaved as designed. Missing, invalid or expired authorization
released no evidence. A one-byte CPE queue dropped 12 records (loss rate 0.869)
but retained 11 verified items. Content tampering quarantined one item; custody-
link tampering quarantined all 17 and left claims unsupported. Home Assistant
unavailability removed the automation claim while entry and tamper remained
strong. Maximum clock skew widened maximum uncertainty from 4.43~s to 60.43~s.
All 16 branch assertions and 47 final tests passed.

Across 100 runs, mean capture recall was 0.899 (95\% interval 0.839--0.959,
$n=99$), mean reconstruction recall 0.665 (0.605--0.725), mean strong-claim count
2.03 (1.83--2.23), and zero-unauthorized-disclosure occurred in 100/100 runs
(Wilson 95\% interval 0.963--1.0). Mean twin lag was 4.87~s; its strongest
reported linear sensitivity was clock-offset bound ($r=0.883$). Collection
start delay was nearly perfectly correlated with collection and investigation
latency in the declared design, a structural consequence rather than a causal
field finding.

\textbf{Framework-validity interpretation.} UC4 is the most complete demonstration of the framework.
It shows that authorization, minimization and tamper detection can be tested as
executable invariants and that useful claims can survive partial source loss.
Its reconstruction precision, stale-state fraction and delay sensitivities also
show why a successful nominal narrative must not be mistaken for calibration.

\subsubsection{UC5: Clinical Continuity Before Isolation}

\textbf{Simulation design.} The versioned 60-s configurations define an essential patient
monitor with no nominal backup, 20 normal flows of 1,000 bytes, a symbolic
fourfold anomalous-egress multiplier and a new destination. The detector weights
rate/bytes/destination/health as 0.25/0.25/0.30/0.20 at threshold 0.5. Isolation
is requested but must be safety-rejected without backup. Nine configured branches
cover baseline, nominal, flood, observation loss, backup state, enforcement
failure, health fault and sharing denial; a 100-sample design varies attack
multiplier, observation/loss and threshold.

\textbf{Results.} The nominal run generated and observed 80 abstract flows,
raised one explainable alert and requested isolation. The safety gate rejected
that request because the essential monitor had no approved backup; consequently
process effects and essential-monitoring interruptions were zero. It verified
and shared 84 minimized evidence items with zero patient-data disclosures,
privacy violations, safety violations or integrity failures. In the
backup-active branch, one bounded isolation was eligible while interruption
remained zero. All nine branches passed; unauthorized sharing exported no
evidence IDs and a health-only fault produced no cyber alert. Across 100 runs,
mean alert and safety-rejection counts were both 1.0, with zero essential
interruptions, patient disclosures, privacy violations and safety violations.
Thirteen retained tests passed.

\textbf{Framework-validity interpretation.} UC5 makes clinical continuity an executable hard
constraint and demonstrates fail-closed evidence sharing. These synthetic
invariants do not establish clinical safety, hospital fidelity or regulatory
compliance.

\subsection{Framework-Validity Findings Across Use Cases}

Tables~\ref{tab:nominal-results}--\ref{tab:validation-results} summarize the
simulation evidence by experiment type. In each table, the use cases are rows
and the cells contain only numeric outcomes. Blank cells denote metrics that are
not defined by the corresponding simulator. The column headings and captions
define the metric, while the preceding use-case text explains the
domain-specific meaning of each number.

\begin{table}[ht]
\caption{Nominal experiment results. ``Gen.'' and ``Obs.'' count generated and
observed messages, observations, flows or evidence-source records; ``Evid.''
is evidence-capture recall for UC1 and preserved/shared evidence items for UC4
and UC5; ``Rec.'' and ``Prec.'' are reconstruction recall and precision.}
\label{tab:nominal-results}
\centering
\scriptsize
\setlength{\tabcolsep}{3pt}
\begin{tabular}{lrrrrrrrrr}
\toprule
UC & Gen. & Obs. & Lost & Alerts & Evid. & Rec. & Prec. & Effects & Viol. \\
\midrule
UC1 & 3 & 3 & 0 & 3 & 1.0 &  &  & 0 & 0 \\
UC2 & 48 & 45 & 3 & 1 &  &  &  & 0 & 0 \\
UC3 & 6 & 6 & 0 & 6 &  &  &  & 0 & 0 \\
UC4 & 17 & 17 & 0 &  & 17 & 1.0 & 0.5 &  & 0 \\
UC5 & 80 & 80 & 0 & 1 & 84 &  &  & 0 & 0 \\
\bottomrule
\end{tabular}
\end{table}

\begin{table}[ht]
\caption{Stress and degraded-branch experiment results. ``Pass'' counts
passing branches; ``Overflow'' and ``ObsLoss'' report the maximum observed loss
counts in overflow and observation-loss branches; ``LossAlerts'' counts alerts
when observations are completely lost; ``Blocks'' counts explicit authority or
safety blocking decisions; ``Quar.'' counts quarantined evidence items; and
``Effects'' reports the maximum simulated process effect count.}
\label{tab:stress-results}
\centering
\scriptsize
\setlength{\tabcolsep}{2pt}
\begin{tabular}{lrrrrrrrr}
\toprule
UC & Branches & Pass & Overflow & ObsLoss & LossAlerts & Blocks & Quar. &
Effects \\
\midrule
UC1 & 9 & 9 & 47 & 6 & 0 & 6 & 0 & 0 \\
UC2 & 9 & 9 & 46 & 48 & 0 & 0 & 0 & 0 \\
UC3 & 9 & 9 & 0 &  & 0 & 1 & 0 & 0 \\
UC4 & 16 & 16 & 12 & 1 &  & 1 & 17 & 0 \\
UC5 & 9 & 9 & 100 & 80 & 0 & 1 & 0 & 1 \\
\bottomrule
\end{tabular}
\end{table}

\begin{table}[ht]
\caption{Uncertainty experiment results. Values are Monte Carlo means over 100
predeclared simulation runs per use case.
``DetRec'' denotes attack-detection recall, ``CapRec'' evidence-capture recall,
``ReconRec'' reconstruction recall, ``Claims'' strong reconstructed claims,
``Loss'' observation loss and ``Reject'' safety rejections.}
\label{tab:uncertainty-results}
\centering
\scriptsize
\setlength{\tabcolsep}{3pt}
\begin{tabular}{lrrrrrrrr}
\toprule
UC & Alerts & DetRec & CapRec & ReconRec & Claims & Loss & Reject &
Viol. \\
\midrule
UC1 &  & 0.28 & 1.0 &  &  &  &  & 0 \\
UC2 & 0.60 &  &  &  &  & 13.07 &  & 0 \\
UC3 & 3.88 &  &  &  &  &  &  & 0 \\
UC4 &  &  & 0.899 & 0.665 & 2.03 &  &  & 0 \\
UC5 & 1.0 &  &  &  &  &  & 1.0 & 0 \\
\bottomrule
\end{tabular}
\end{table}

\begin{table}[!ht]
\caption{Validation and implementation evidence. All listed packages are
internally validated against the same artifact contract.}
\label{tab:validation-results}
\centering
\scriptsize
\setlength{\tabcolsep}{4pt}
\begin{tabular}{lrrrrrr}
\toprule
UC & Tests & Branches & Entities & Relations & Schemas & BPMN \\
\midrule
UC1 & 22 & 9 & 30 & 60 & 9 & 7 \\
UC2 & 11 & 9 & 25 & 26 & 9 & 7 \\
UC3 & 21 & 9 & 36 & 60 & 9 & 8 \\
UC4 & 47 & 16 & 38 & 123 & 18 & 7 \\
UC5 & 13 & 9 & 34 & 45 & 9 & 8 \\
\bottomrule
\end{tabular}
\end{table}

Table~\ref{tab:crossresults} summarizes the main result and supported claim for
each use case.

\begin{table*}[t]
\caption{Summary of simulation results and framework-validity interpretation
across the use cases.}
\label{tab:crossresults}
\centering
\small
\begin{tabularx}{\textwidth}{L{12mm}L{29mm}YYY}
\toprule
UC & Primary decision & Nominal observable result & Stress evidence & Defensible claim \\
\midrule
UC1 & Trust and safety before control & 3/3 attacks alerted; no controller effect
& Flood recall 0.06; loss gives no alerts & Fail-safe path works in synthetic model \\
UC2 & Passive, pseudonymous observation & 45 observations; 4 classes; 1 alert
& Loss/overflow and stale dependencies fail closed & Passive/privacy invariants hold \\
UC3 & Safety guard before mitigation & 6 alerts; no process effect & Loss gives no
alerts; all branches pass & Surrogate invariants hold \\
UC4 & Authority and minimization before release & 17 verified items; three
strong claims; zero disclosure & Tamper quarantined; source loss degrades claims
& Full synthetic assurance chain works \\
UC5 & Care continuity before isolation & 1 alert; isolation rejected; no interruption
& Backup permits bounded isolation; unauthorized share denied & Continuity/privacy invariants hold \\
\bottomrule
\end{tabularx}
\end{table*}

The results support the framework's internal validity in four ways. First, the
cases could be represented through the same source-governed case,
canonical model, process/behaviour, security-governance and assurance-package
structure. This supports wider applicability, because the cases differ in
referents, observations, policies, evidence rules and safety constraints.
Second, the simulations made failed or degraded conditions observable rather
than hiding them: observation loss removed alerts, tampering quarantined
evidence, unsafe isolation was rejected and source loss weakened claims. Third,
zero harmful effects were meaningful only where attempted transitions and guard
rejections were also recorded; otherwise a scenario could simply be inert.
Fourth, Eq.~\eqref{eq:policy} was reusable as a governance pattern, but the
meaning of safety or authority remained domain-specific: trust in UC1, passive
privacy in UC2, process safety in UC3, authorization in UC4 and clinical
continuity in UC5 required different expert-owned predicates.

The same results also bound the validity claim. They validate the framework as a
reproducible method for building internally consistent, evidence-qualified
security-DT simulations. They do not validate operational performance, detector
accuracy, legal admissibility, clinical safety or industrial fidelity. Those
stronger claims require the future calibration and external validation steps
listed in Section~\ref{sec:limitations}.

\subsection{Counterfactual Assurance Checks}
\label{sec:naive-ablations}

As a final negative check, we ran three deliberately naive variants against the
same five generated packages. The first variant lets the twin read hidden attack
or incident truth even when the observation layer does not contain it. The
second treats a requested security action as if it had also been authorized and
applied. The third removes claim qualification, so that strong claims are not
downgraded after evidence loss, tampering or source degradation. These
ablations are not proposed alternatives; they are controlled anti-patterns used
to test whether the framework rules prevent invalid certainty.

\begin{table*}[t]
\caption{Naive-twin ablation outcomes. Each cell reports what the corresponding
naive variant would claim or do after one \framework{} discipline is removed.}
\label{tab:naive-ablation-results}
\centering
\scriptsize
\setlength{\tabcolsep}{3pt}
\begin{tabularx}{\textwidth}{L{9mm}YYY}
\toprule
UC & Oracle-leaking twin & Requested = applied & No claim qualification \\
\midrule
UC1 & Observation-loss branch changes from 0 observed attacks and 0 alerts to
confident recall 1.0. & Failed enforcement is recorded as applied, overstating
mitigation success. & A forged-message claim remains strong despite missing
attack observations. \\
UC2 & Zero passive observations still produce an unapproved-device claim. &
Passive monitoring invents an active applied response. & A strong
classification survives a two-observation overflow record. \\
UC3 & Missing observations still produce 6 alerts and recall 1.0. & Unsafe
isolation becomes an applied process action, yielding a safety violation. & A
strong attack claim survives no observable attack evidence. \\
UC4 & Source-record eviction still reconstructs a complete incident narrative. &
An authorization request is treated as a release grant. & Three strong claims
survive custody tampering or total source-record loss. \\
UC5 & Zero observed flows still produce a confident compromised-monitor alert. &
Isolation is applied without backup, causing a continuity violation. & A strong
compromise claim survives missing clinical flow evidence. \\
\bottomrule
\end{tabularx}
\end{table*}

Table~\ref{tab:naive-ablation-results} shows the expected pattern. The
oracle-leaking twin looks better than the framework in the wrong places: UC1,
UC3 and UC5 recover high-confidence attacks under complete observation loss,
while UC2 identifies an unapproved device with no passive observations. The
requested-equals-applied twin fails where action semantics matter. In UC3 it
turns an unsafe isolation request into an applied process action, and in UC5 it
isolates an essential monitor without an approved backup. UC4 shows the same
problem in evidence-governance form: a request for access is treated as if it
were an authorization grant. The unqualified-claim twin keeps strong claims
after the evidence base has degraded; the UC4 case is the clearest example,
because strong entry, automation and tamper claims survive custody tampering or
total source-record loss.

These ablations do not measure detector performance. They test whether
attractive but invalid conclusions are blocked by the framework. The answer is
positive in all five packages. Hidden truth is not allowed to become observed
evidence, requested mitigations are not allowed to become applied actions
without a guard, and claims are not allowed to remain strong when their evidence
support has been damaged. This is the same discipline exercised in the nominal
and degraded experiments, but viewed from the opposite direction: by removing
the discipline, the model immediately produces overconfident alerts, unsafe or
unauthorized actions and unsupported claims.

\section{Limitations and Future Work}
\label{sec:limitations}

The evaluation shows that the framework can be implemented consistently across
different security-DT scenarios, but the evidence remains deliberately
bounded. The simulations test internal coherence, traceability, visibility
control, policy enforcement and evidence handling under declared assumptions.
They are not intended to stand in for field measurements, certified safety
cases, legal admissibility assessments or deployment KPIs. The main limitations
therefore concern the distance between an internally validated synthetic package
and an externally calibrated operational system.

The most important limitations are the following.

\begin{itemize}
  \item \textbf{Synthetic and uncalibrated dynamics.} The parameters used in the
  experiments are explicit simulation assumptions. The reported values should
  therefore be read as evidence about the framework implementation, not as
  estimates of operational performance in the corresponding real environments.
  \item \textbf{Uneven implementation depth.} Each package satisfies the common
  internal completion contract. However, UC1--UC3 and UC5 use compact
  formal models, whereas UC4 contains deeper equations, experiments and
  assurance artefacts. This limits the breadth of cross-UC comparison without
  changing the completion status of the individual packages.
  \item \textbf{Simplified detection models.} The detectors use transparent
  weighted rules rather than trained or externally validated AI models. The
  Monte Carlo ranges capture designer-specified uncertainty, not learned
  parameter distributions.
  \item \textbf{Limited physical fidelity.} The network models use event and
  flow queues rather than packet-level or RF emulation. Controllers and sensors
  are bounded surrogates, and the simulations do not include geometry, protocol
  stack behaviour or hardware timing.
  \item \textbf{Bounded evidence claims.} Hashes and custody chains detect
  modelled alteration within the package. They do not authenticate external
  collectors or keys, prove legal authority, or establish legal admissibility.
  \item \textbf{Limited metric comparability.} Similar metric labels can refer
  to different domain populations. Cross-UC aggregation would require a common
  measurement ontology, shared denominators and agreed reporting conventions.
  \item \textbf{No independent validation yet.} The tests were produced within
  the project and focus on implementation. External datasets, testbeds, domain
  experts and red-team review are needed to challenge the model structure.
  \item \textbf{Cross-package drift risk.} Keeping per-UC packages protects
  domain distinctions, but it also duplicates schemas and utilities. Premature
  consolidation would hide important differences; leaving everything separate
  indefinitely would allow divergence.
\end{itemize}

\subsection{Prioritized future work}

The next phase should move from implementation closure to expert calibration
and independent evaluation. The first step is to convene use-case expert reviews
that resolve the open parameters listed in the scenario descriptions. Each
accepted value should be recorded with its source, unit, uncertainty, owner and
effective version. Candidate datasets should then be introduced only after their
licence, checksum, field mappings, time mappings, identity mappings and
missingness assumptions are approved. The synthetic modes should remain as
reproducible controls.

A second line of work is architectural. The project should define a minimal
common kernel for stable identifiers, event envelopes, visibility labels,
canonical hashing, custody, null-aware metrics and manifests, while keeping
physical dynamics, observation models, policy predicates and domain schemas in
the per-UC packages. Canonical entities and provenance records should then be
mapped to interoperable representations such as AAS and W3C PROV, with
round-trip loss measured explicitly.

A third line of work is validation. Mutation and property tests, model checking
of critical invariants, differential checks between equations and code,
independent security review and continuous manifest verification would make the
assurance packages harder to satisfy accidentally. After that, calibration,
hardware/software-in-the-loop experiments and pilot validation can be introduced
under a staged claim protocol: calibrated results must identify their dataset
and uncertainty, while deployed results must identify their adapter,
authorization path and safety case. Operational KPIs should be reported only
after denominators, baselines and validation protocols have been agreed.

\section{Conclusions}
\label{sec:conclusions}

The primary outcome is a framework for making cyber-physical Digital Twin work
auditable from definition to result. \framework{} treats an entity graph,
mathematical model, process diagram, simulator, evidence record and test report
as mutually constraining views of one system. Its proposed DT definition adds
observation admissibility, fidelity, freshness, provenance and authority to the
usual physical--digital correspondence. Its central invariant prevents hidden
simulation truth from silently entering operational state or forensic claims.

The evaluation indicates that this structure can be applied beyond a single
case without reducing every system to the same model. The important result is
not that the simulations reproduce operational environments, but that they
preserve the same assurance discipline while allowing different assets,
observations, policies, safety constraints and evidence rules. In each case,
normal and degraded runs remain traceable to declared assumptions, and failed
conditions become explicit results rather than silent gaps in the model.

The broader conclusion is that secure DTs need an assurance architecture, not
only a synchronized state model. A useful DT for security analysis must say what
it can observe, what it cannot observe, which actions it may support, which
claims are evidence-backed and where uncertainty remains. \framework{} provides
a practical way to encode those boundaries so that simulations, tests and
reports can be reviewed together.

The framework's value is therefore not a promise of universal fidelity or
deployment readiness. It is a disciplined way to state what is known, what is
assumed, what was observed, what was decided and what evidence supports each
claim. That discipline is a necessary foundation for later interoperability,
expert calibration, independent validation and responsible deployment of
security-oriented Digital Twin systems.

\section*{Acknowledgment}

This work has received funding from the European Union's Horizon Europe
research and innovation programme under Grant Agreement No. 101225719 (SALUS:
Strengthening law enforcement with advanced IoT forensic tools and enriched
investigation schemes, realised by a Software Defined Network
Security-as-a-Service architecture).
Views and opinions expressed are however
those of the author(s) only and do not necessarily reflect those of the
European Union or the granting authority. Neither the European Union nor the
granting authority can be held responsible for them.

\appendix
\section{UC1 Car2X Implementation Detail}
\label{sec:appendix-uc1}

This appendix records the implementation-level content behind the UC1
Car2X evaluation. The purpose is not to claim operational Car2X fidelity, but
to make the synthetic implementation inspectable: the reader can see which
entities exist, which relationships bind them, which assumptions remain, and
which experiment branches were used to test the framework. These details are reported in
Tables~\ref{tab:uc1-entities}--\ref{tab:uc1-validation}.


\begin{thebibliography}{99}
\bibitem{glaessgen2012}
E. H. Glaessgen and D. S. Stargel,
``The Digital Twin Paradigm for Future NASA and U.S. Air Force Vehicles,''
53rd AIAA/ASME/ASCE/AHS/ASC Structures, Structural Dynamics and Materials
Conference, AIAA 2012-1818, 2012. doi: \href{https://doi.org/10.2514/6.2012-1818}{10.2514/6.2012-1818}.

\bibitem{iso23247}
International Organization for Standardization,
\emph{ISO 23247-1:2021, Automation systems and integration---Digital twin
framework for manufacturing---Part 1: Overview and general principles}, 2021.
\url{https://www.iso.org/standard/75066.html}.

\bibitem{dtcinterop}
Digital Twin Consortium,
\emph{Digital Twin System Interoperability Framework}, 2021.
\url{https://www.digitaltwinconsortium.org/wp-content/uploads/sites/3/2022/06/Digital-Twin-System-Interoperability-Framework-12072021.pdf}.

\bibitem{aas}
Industrial Digital Twin Association,
\emph{Specification of the Asset Administration Shell, Part 1: Metamodel},
IDTA 01001, version 3 series.
\url{https://industrialdigitaltwin.org/en/content-hub/aasspecifications}.

\bibitem{nistcps}
E. R. Griffor, C. Greer, D. A. Wollman, and M. J. Burns,
\emph{Framework for Cyber-Physical Systems: Volume 2, Working Group Reports},
NIST SP 1500-202, 2017. doi: \href{https://doi.org/10.6028/NIST.SP.1500-202}{10.6028/NIST.SP.1500-202}.

\bibitem{yu2023cpssecurity}
Z. Yu, H. Gao, X. Cong, N. Wu, and H. H. Song,
``A Survey on Cyber-Physical Systems Security,'' \emph{IEEE Internet of Things
Journal}, vol. 10, no. 24, pp. 21670--21686, 2023. doi: \href{https://doi.org/10.1109/JIOT.2023.3289625}{10.1109/JIOT.2023.3289625}.

\bibitem{jeremiah2024dtsurvey}
S. R. Jeremiah, A. El Azzaoui, N. N. Xiong, and J. H. Park,
``A Comprehensive Survey of Digital Twins: Applications, Technologies and
Security Challenges,'' \emph{Journal of Systems Architecture}, vol. 151,
Art. no. 103120, 2024. doi: \href{https://doi.org/10.1016/j.sysarc.2024.103120}{10.1016/j.sysarc.2024.103120}.

\bibitem{qureshi2025securitydt}
A. R. Qureshi, A. Asensio, M. Imran, J. Garcia, and X. Masip-Bruin,
``A Survey on Security Enhancing Digital Twins: Models, Applications and
Tools,'' \emph{Computer Communications}, vol. 238, Art. no. 108158, 2025.
doi: \href{https://doi.org/10.1016/j.comcom.2025.108158}{10.1016/j.comcom.2025.108158}.

\bibitem{balta2023}
E. Balta, M. Pease, J. Moyne, K. Barton, and D. Tilbury,
``Digital Twin-Based Cyber-Attack Detection Framework for Cyber-Physical
Manufacturing Systems,'' \emph{IEEE Transactions on Automation Science and
Engineering}, 2023. doi: \href{https://doi.org/10.1109/TASE.2023.3243147}{10.1109/TASE.2023.3243147}.

\bibitem{moradi2024crystal}
F. Moradi, S. Abbaspour Asadollah, B. Pourvatan, Z. Moezkarimi, and
M. Sirjani, ``CRYSTAL Framework: Cybersecurity Assurance for Cyber-Physical
Systems,'' \emph{Journal of Logical and Algebraic Methods in Programming},
vol. 139, Art. no. 100965, 2024. doi: \href{https://doi.org/10.1016/j.jlamp.2024.100965}{10.1016/j.jlamp.2024.100965}.

\bibitem{eckhart2019cpssecurity}
M. Eckhart and A. Ekelhart,
``Digital Twins for Cyber-Physical Systems Security: State of the Art and
Outlook,'' in \emph{Security and Quality in Cyber-Physical Systems Engineering},
Springer, 2019, pp. 383--412. doi: \href{https://doi.org/10.1007/978-3-030-25312-7_14}{10.1007/978-3-030-25312-7\_14}.

\bibitem{dietz2020soc}
M. Dietz, M. Vielberth, and G. Pernul,
``Integrating Digital Twin Security Simulations in the Security Operations
Center,'' in \emph{Proceedings of the 15th International Conference on
Availability, Reliability and Security (ARES 2020)}, ACM, 2020. doi: \href{https://doi.org/10.1145/3407023.3407039}{10.1145/3407023.3407039}.

\bibitem{eckhart2023sedt}
M. Eckhart, A. Ekelhart, D. Allison, M. Almgren, K. Ceesay-Seitz, H. Janicke,
S. Nadjm-Tehrani, A. Rashid, and M. Yampolskiy,
``Security-Enhancing Digital Twins: Characteristics, Indicators, and Future
Perspectives,'' arXiv:2305.00639, 2023.
\url{https://arxiv.org/abs/2305.00639}.

\bibitem{katsikas2026industrialdt}
K. E. Kampourakis, V. Gkioulos, and S. Katsikas,
``Cybersecurity Digital Twins for Industrial Systems: From Literature Synthesis
to Framework Design,'' \emph{Information}, vol. 17, no. 3, Art. no. 286, 2026.
doi: \href{https://doi.org/10.3390/info17030286}{10.3390/info17030286}.

\bibitem{dietz2021icsforensics}
M. Dietz, L. Englbrecht, and G. Pernul,
``Enhancing Industrial Control System Forensics Using Replication-Based Digital
Twins,'' in \emph{Advances in Digital Forensics XVII}, IFIP Advances in
Information and Communication Technology, vol. 612, Springer, 2021,
pp. 21--38. doi: \href{https://doi.org/10.1007/978-3-030-88381-2_2}{10.1007/978-3-030-88381-2\_2}.

\bibitem{katsikas2026attckdt}
K. E. Kampourakis, V. Kampourakis, V. Gkioulos, and S. Katsikas,
``Digital Twin-Assisted Mapping of ICS Telemetry to ATT\&CK for ICS with
Evidence-Driven Dependency Reasoning,'' arXiv:2610.02955, 2026.
\url{https://arxiv.org/abs/2610.02955}.

\bibitem{nistdtcredibility}
Y. Lu, S. Y. Liang, and K. Morris,
``Credibility Consideration for Digital Twins in Manufacturing,'' NIST,
2023. doi: \href{https://doi.org/10.6028/NIST.AMS.100-49}{10.6028/NIST.AMS.100-49}.

\bibitem{asmevv40}
ASME,
\emph{V\&V 40-2018: Assessing Credibility of Computational Modeling Through
Verification and Validation: Application to Medical Devices}, 2018.
\url{https://www.asme.org/codes-standards}.

\bibitem{nasastd7009}
National Aeronautics and Space Administration,
\emph{NASA-STD-7009B: Standard for Models and Simulations}, 2024.
\url{https://standards.nasa.gov/standard/nasa/nasa-std-7009}.

\bibitem{sha1996simplex}
J. G. Rivera, A. A. Danylyszyn, C. B. Weinstock, L. Sha, and M. J. Gagliardi,
\emph{An Architectural Description of the Simplex Architecture},
CMU/SEI-96-TR-006, Software Engineering Institute, Carnegie Mellon University,
1996.
\url{https://www.sei.cmu.edu/library/an-architectural-description-of-the-simplex-architecture/}.

\bibitem{bloem2015shield}
R. Bloem, B. K\"onighofer, R. K\"onighofer, and C. Wang,
``Shield Synthesis: Runtime Enforcement for Reactive Systems,'' in
\emph{Tools and Algorithms for the Construction and Analysis of Systems},
Springer, 2015, pp. 533--548. doi: \href{https://doi.org/10.1007/978-3-662-46681-0_51}{10.1007/978-3-662-46681-0\_51}.

\bibitem{nistir2025}
A. Nelson, S. Rekhi, M. Souppaya, and K. Scarfone,
\emph{Incident Response Recommendations and Considerations for Cybersecurity
Risk Management: A CSF 2.0 Community Profile}, NIST SP 800-61 Rev. 3, 2025.
doi: \href{https://doi.org/10.6028/NIST.SP.800-61r3}{10.6028/NIST.SP.800-61r3}.

\bibitem{oscal}
National Institute of Standards and Technology,
\emph{OSCAL: Open Security Controls Assessment Language}, 2026.
\url{https://pages.nist.gov/OSCAL/}.

\bibitem{ocsf}
Open Cybersecurity Schema Framework,
\emph{OCSF Schema}, version 1.9.0, 2026.
\url{https://github.com/ocsf/ocsf-schema/releases/tag/1.9.0}.

\bibitem{stix21}
OASIS Cyber Threat Intelligence Technical Committee,
\emph{STIX Version 2.1}, OASIS Committee Specification 01, 2021.
\url{https://docs.oasis-open.org/cti/stix/v2.1/stix-v2.1.html}.

\bibitem{taxii21}
OASIS Cyber Threat Intelligence Technical Committee,
\emph{TAXII Version 2.1}, OASIS Committee Specification 01, 2021.
\url{https://docs.oasis-open.org/cti/taxii/v2.1/taxii-v2.1.html}.

\bibitem{csaf20}
OASIS Common Security Advisory Framework Technical Committee,
\emph{Common Security Advisory Framework Version 2.0}, OASIS Standard, 2022.
\url{https://docs.oasis-open.org/csaf/csaf/v2.0/os/csaf-v2.0-os.html}.

\bibitem{jsonschema}
A. Wright, H. Andrews, B. Hutton, and G. Dennis,
\emph{JSON Schema: A Media Type for Describing JSON Documents, Draft 2020-12},
2022. \url{https://json-schema.org/draft/2020-12/json-schema-core.html}.

\bibitem{prov}
World Wide Web Consortium,
\emph{PROV-DM: The PROV Data Model}, W3C Recommendation, 30 April 2013.
\url{https://www.w3.org/TR/prov-dm/}.

\bibitem{nistforensics}
K. Kent, S. Chevalier, T. Grance, and H. Dang,
\emph{Guide to Integrating Forensic Techniques into Incident Response},
NIST SP 800-86, 2006. doi: \href{https://doi.org/10.6028/NIST.SP.800-86}{10.6028/NIST.SP.800-86}.

\bibitem{bpmn}
Object Management Group,
\emph{Business Process Model and Notation (BPMN), Version 2.0.2}, 2014.
\url{https://www.omg.org/spec/BPMN/2.0.2/PDF/}.

\bibitem{gsnstandard}
Assurance Case Working Group,
\emph{Goal Structuring Notation Community Standard, Version 3}, Safety-Critical
Systems Club, 2021.
\url{https://scsc.uk/scsc-141B}.

\bibitem{sacm20}
Object Management Group,
\emph{Structured Assurance Case Metamodel (SACM), Version 2.0}, 2018.
\url{https://www.omg.org/spec/SACM/2.0/PDF}.

\bibitem{iso27037}
International Organization for Standardization and International
Electrotechnical Commission,
\emph{ISO/IEC 27037:2012, Information technology---Security techniques---
Guidelines for identification, collection, acquisition and preservation of
digital evidence}, 2012.
\url{https://www.iso.org/standard/44381.html}.

\bibitem{iso27043}
International Organization for Standardization and International
Electrotechnical Commission,
\emph{ISO/IEC 27043:2015, Information technology---Security techniques---
Incident investigation principles and processes}, 2015.
\url{https://www.iso.org/standard/44407.html}.

\bibitem{zhao2022dtcyber}
T. Zhao, E. Foo, and H. Tian,
``A Digital Twin Framework for Cyber Security in Cyber-Physical Systems,''
arXiv:2204.13859, 2022.
\url{https://arxiv.org/abs/2204.13859}.

\bibitem{suhail2022trustedtwins}
S. Suhail, S. U. R. Malik, R. Jurdak, R. Hussain, R. Matulevi\v{c}ius, and
D. Svetinovic,
``Towards Situational Aware Cyber-Physical Systems: A Security-Enhancing Use
Case of Blockchain-Based Digital Twins,'' \emph{Computers in Industry},
vol. 141, Art. no. 103699, 2022. doi: \href{https://doi.org/10.1016/j.compind.2022.103699}{10.1016/j.compind.2022.103699}.

\bibitem{suhail2025sedtir}
S. Suhail, M. Iqbal, K. McLaughlin, B. Lee, and B. Imtiaz,
``A Framework for Enhancing Cyber Incident Response with Security-Enhancing
Digital Twins in Cyber--Physical Systems,'' \emph{Internet of Things},
vol. 31, Art. no. 101547, 2025. doi: \href{https://doi.org/10.1016/j.iot.2025.101547}{10.1016/j.iot.2025.101547}.

\bibitem{repetto2026cdt}
M. Repetto,
``Cybersecurity Digital Twins: Concept, Blueprint, and Challenges for
Multi-Ownership Digital Service Chains,'' \emph{Journal of Information
Security and Applications}, vol. 96, Art. no. 104299, 2026. doi: \href{https://doi.org/10.1016/j.jisa.2025.104299}{10.1016/j.jisa.2025.104299}.

\end{thebibliography}
\end{document}